\documentclass[conference]{IEEEtran}
\IEEEoverridecommandlockouts
\usepackage{subcaption}
\usepackage{listings}
\usepackage{xcolor}
\usepackage{todonotes}
\usepackage{amsmath}
\usepackage{pifont}
\usepackage{graphicx}
\usepackage[shortlabels]{enumitem}
\usepackage{color}
\usepackage{tikz}
\usepackage{balance}
\usepackage{multirow}
\usepackage{comment}
\usepackage{amsmath}
\usepackage{graphicx}
\usepackage{wrapfig,lipsum,booktabs}
\usepackage{titlesec}
\usepackage{soul}
\usepackage{placeins}

\usepackage{dblfloatfix}

\definecolor{codegreen}{rgb}{0,0.6,0}
\definecolor{codegray}{rgb}{0.5,0.5,0.5}
\definecolor{codepurple}{rgb}{0.58,0,0.82}
\definecolor{backcolour}{rgb}{0.95,0.95,0.92}

\lstdefinestyle{mystyle}{
  backgroundcolor=\color{backcolour},   commentstyle=\color{codegreen},
  keywordstyle=\color{magenta},
  numberstyle=\tiny\color{codegray},
  stringstyle=\color{codepurple},
  basicstyle=\ttfamily\footnotesize,
  breakatwhitespace=false,         
  breaklines=true,                 
  captionpos=b,                    
  keepspaces=true,                 
  numbers=left,                    
  numbersep=5pt,                  
  showspaces=false,                
  showstringspaces=false,
  showtabs=false,                  
  tabsize=2,
  xleftmargin=2.0ex
}

\def\BibTeX{{\rm B\kern-.05em{\sc i\kern-.025em b}\kern-.08em
    T\kern-.1667em\lower.7ex\hbox{E}\kern-.125emX}}

\begin{document}
\title{Demystifying the Communication Characteristics for Distributed Transformer Models}


\author{\IEEEauthorblockN{Quentin Anthony$^*$, Benjamin Michalowicz$^*$, Jacob Hatef, Lang Xu, \\ Mustafa Abduljabbar, Aamir Shafi, Hari Subramoni, Dhabaleswar K. (DK) Panda \thanks{$^*$ denotes equal contribution}}

\IEEEauthorblockA{
                \textit{Department of Computer Science and Engineering},
                \textit{The Ohio State University},
                Columbus, Ohio, USA \\
                \{anthony.301, michalowicz.2, hatef.4, xu.3304, abduljabbar.1, shafi.16, subramoni.1, panda.2\}@osu.edu
                }

}
\maketitle

\begin{abstract}
Deep learning (DL) models based on the transformer architecture have revolutionized many DL applications such as large language models (LLMs), vision transformers, audio generation, and time series prediction. Much of this progress has been fueled by distributed training, yet distributed communication remains a substantial bottleneck to training progress.
This paper examines the communication behavior of transformer models --- that is, how different parallelism schemes used in multi-node/multi-GPU DL Training communicate data in the context of transformers. We use GPT-based language models as a case study of the transformer architecture due to their ubiquity. We validate the empirical results obtained from our communication logs using analytical models. At a high level, our analysis reveals a need to optimize small message point-to-point communication further, correlations between sequence length, per-GPU throughput, model size, and optimizations used,  and where to potentially guide further optimizations in framework and HPC middleware design and optimization.
\end{abstract}
\begin{IEEEkeywords}
Neural Networks, DNN, MPI, GPU, Large Language Models, Interconnects, Communication Characterization
\end{IEEEkeywords}
\vspace{-2ex}
\section{Introduction}
\label{sec:intro}
Large Language Models (LLMs) such as ChatGPT~\cite{openai2024gpt4}, Gemini~\cite{geminiteam2024gemini}, and Llama~\cite{touvron2023llama} are revolutionizing multiple industries with their ability to perform a range of tasks from customer service to creative content generation. LLMs are typically pre-trained with internet-scale, pre-processed data that allows them to learn the intricacies of human languages. After pre-training, LLMs undergo a fine-tuning process in a supervised setting that allows them to excel in downstream tasks like generation, summarization, translation, and question/answering. Modern LLMs utilize a large number of parameters that imply increased computational and memory requirements during training. A higher number of parameters allows the model to capture more intricate relationships and nuances in language, leading to improved performance on a range of downstream tasks.  
\subsection{Motivation}
\vspace{-.3ex}
\label{sec:motivation}
\vspace{-1ex}
\begin{figure}[ht!]
    \centering
    \includegraphics[width=.90\linewidth]{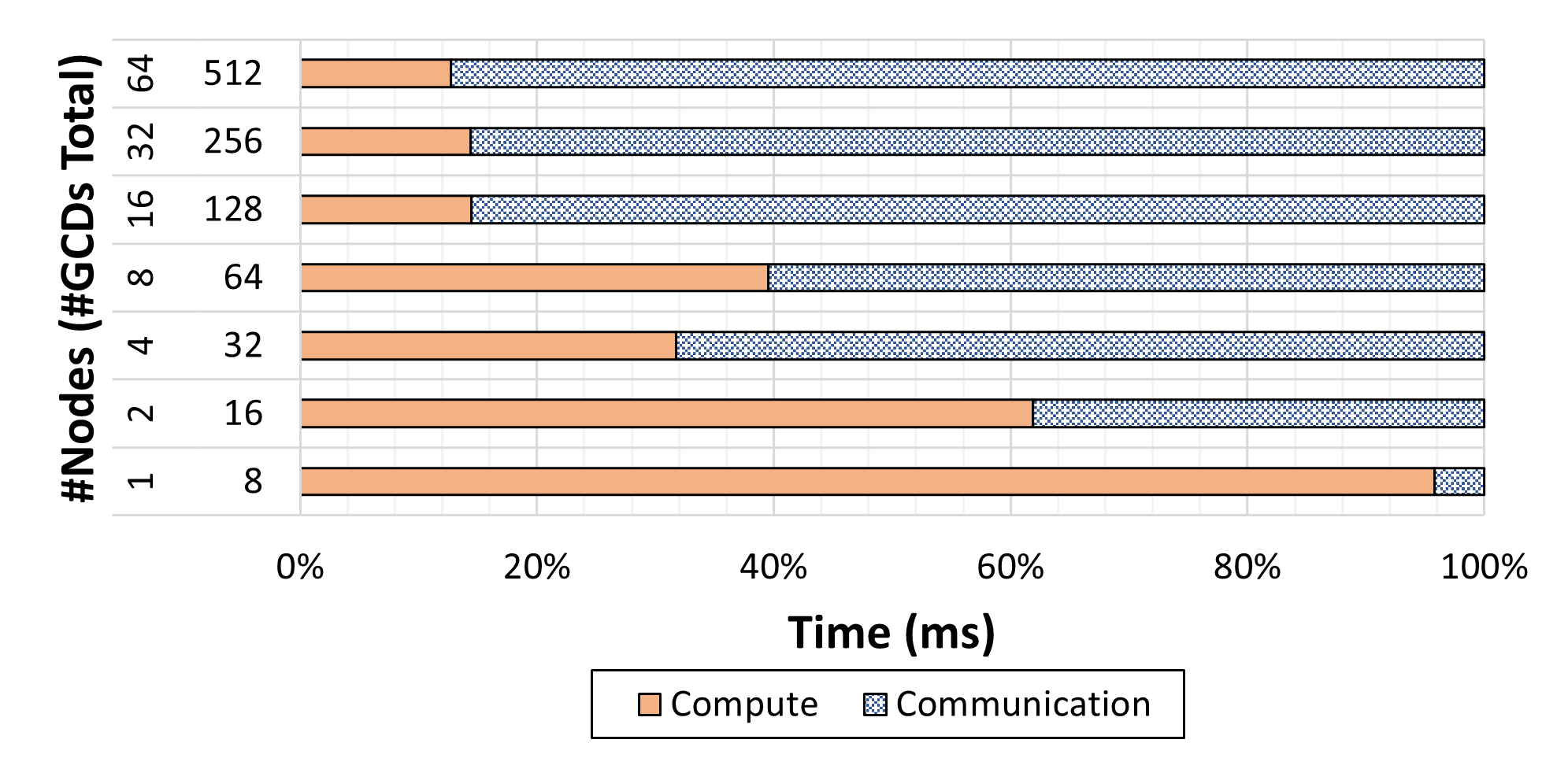}
\vspace{-1ex}
    \caption{13-billion parameter model breakdown of communication and computation using ZeRO-1 and 8 tensor-parallel stages (single iteration)}
    \label{fig:13B_Motivation}
\end{figure}

\begin{figure}[ht!]
    \centering
    \includegraphics[width=.90\linewidth]{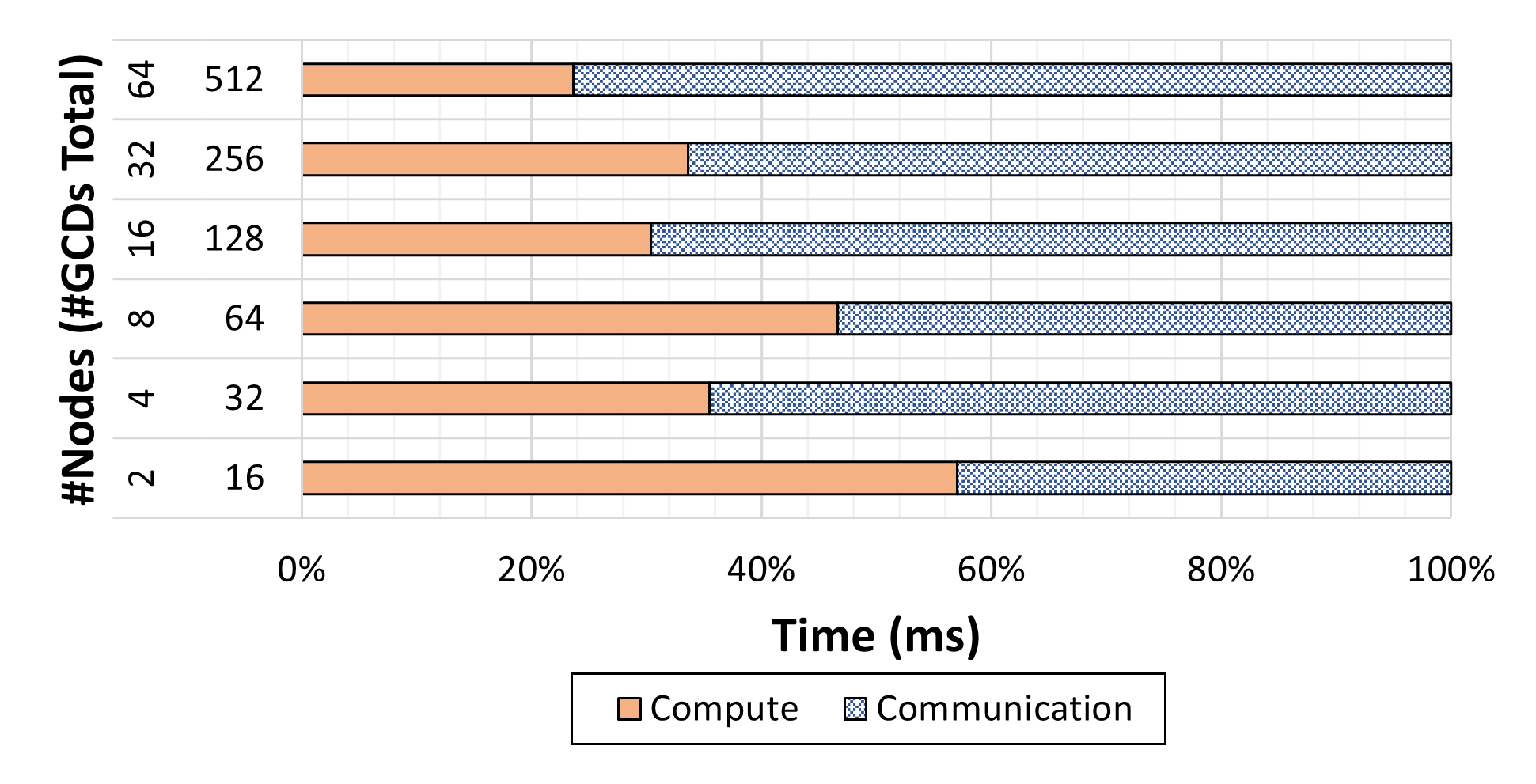}
\vspace{-1ex}
    \caption{20-billion parameter model breakdown of communication and computation using ZeRO-1 and 8 tensor-parallel stages (single iteration)}
    \label{fig:20B_Motivation}
\end{figure}

As an LLM's size increases, training requires a large number of GPUs for a considerable amount of time on modern HPC systems, and it is significantly bottlenecked by how quickly data can be exchanged between parallel training processes. Here, the messaging stack including the communication fabric plays a pivotal role. At large scales, such a bottleneck leads to lower Model FLOPs Utilization (MFU)~\cite{PaLM24} for training. For instance, MegaScale~\cite{MegaScale} reports a 55.2\% MFU on 12,288 GPUs for training a 175-billion parameter model. To emphasize this point, Figures \ref{fig:13B_Motivation} and \ref{fig:20B_Motivation} show how communication begins to dominate computation at increasing scales for 13-billion and 20-billion parameter GPT-2-based models. We are motivated by this to conduct a thorough characterization study to understand the communication stage during LLM training.

\subsection{Problem Statement}
\label{sec:problem-statement}

Good communication performance is critical for scaling LLM training on large HPC systems. This paper aims to study and analyze communication strategies used by state-of-the-art Deep Learning (DL) training frameworks on leading-class supercomputers. Our objective is to learn the volume of data exchanged---as well as communication primitives employed, number of calls, and message sizes involved---between parallel processes at different scales from various parallelization strategies. This detailed analysis needs to be conducted in the context of input datasets, model architectures, and model sizes. This characterization study will aid the next generation of communication runtimes to meet the performance requirements of LLM training workloads and increase the effective utilization of large-scale systems. 

\subsection{Challenges}
\begin{figure}[ht!]
    \centering
    \includegraphics[width=0.75\linewidth]{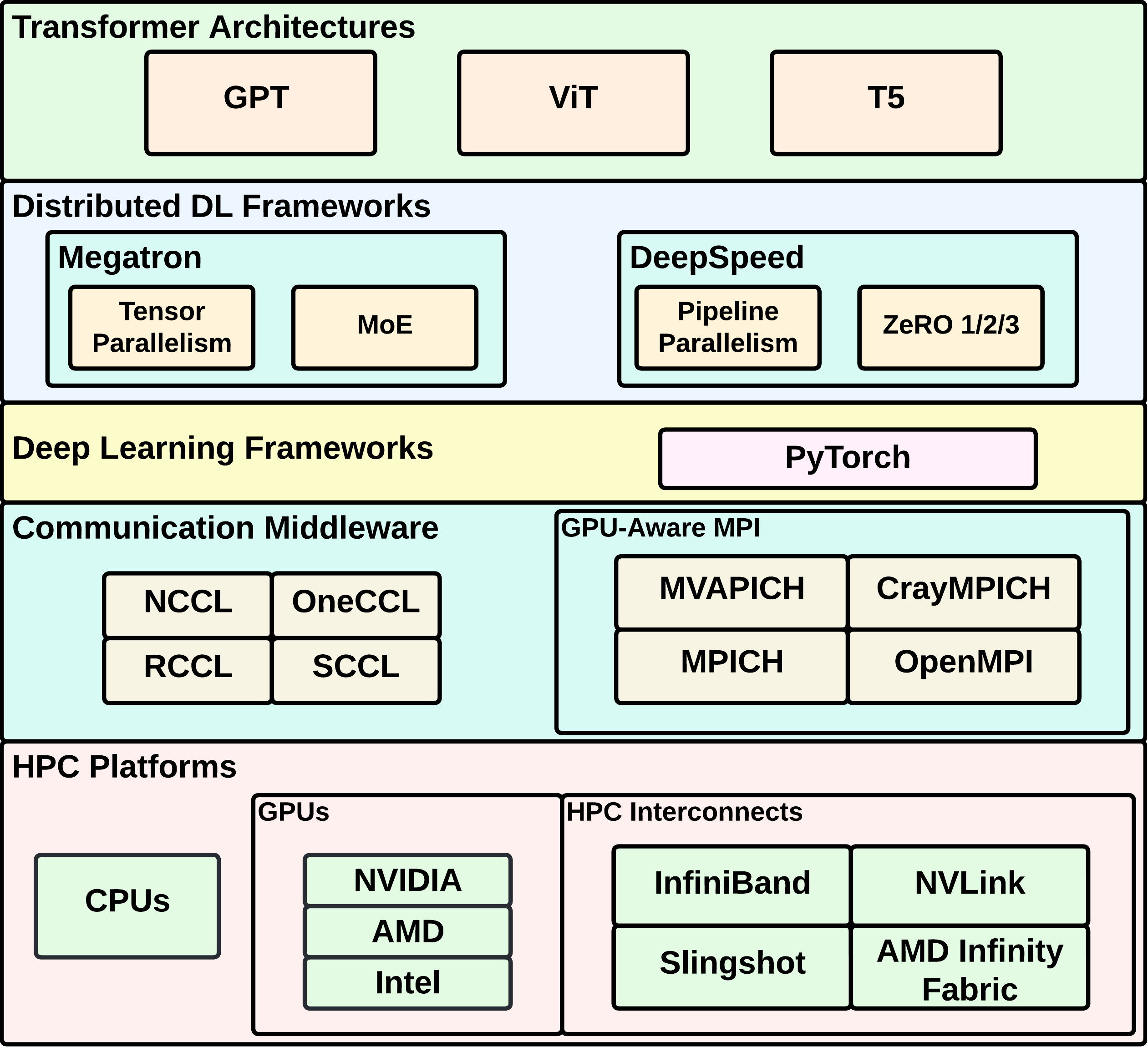}
    \caption{A non-exhaustive list of what must be considered when characterizing LLM performance, scalability, and communication behavior.}
    \label{fig:DL_Combo}
\end{figure}

Figure \ref{fig:DL_Combo} shows just how many combinations someone must consider when characterizing LLM communication on AI/HPC systems, from frameworks such as  Megatron-LM~\cite{megatron-lm}, Llama~\cite{meta-llama}, and DeepSpeed~\cite{deepspeed-mii} and parameter count/model size, to choice of communication middleware \cite{msccl,rccl,nccl}, to parallelism strategies \cite{mcr-dl,awan_hoti_19,Jain-DNN-Char-PyTorch}, all the way down to the hardware on which training/characterization takes place.

Given these challenges, offering insights into communication behavior for transformer architectures while maintaining a balance between the framework, system, and interconnect choices, as well as generality, is not straightforward.

\subsection{Proposed Solution}
\label{sec:proposed-solution}
Given the complexity and importance of understanding communication in emergent transformer-based workloads, we adopt a systematic approach that combines empirical results with analytical modeling to study communication behavior for various parallelism schemes and sequence lengths. Through this, we aim to give an in-depth understanding of the communication overheads associated with parallelism schemes commonly used in transformer models, which form the foundational architecture of LLMs. Our analysis covers a range of model optimizers, including ZeRO-1, ZeRO-2, ZeRO-3, and ZeRO++, as well as Data Parallelism, Pipeline Parallelism, and Tensor Parallelism
for up to 13B parameter models.
In line with the adopted analytical models, we present system-agnostic measurements for each parallelism scheme. Measurements include 1) the collective communication type 2) the data volumes per collective 3) the proportions, frequency, and message sizes for each collective. We also examine the impact of sequence length on communication volumes per collective pattern for Data-Parallel and Model-Parallel environments. This technique is particularly valuable for researchers and developers of collective communication libraries, as it provides insights into which collectives to enhance and which message ranges to target to improve LLM training performance. Additionally, we conduct interconnect-specific evaluations, measuring latency for particular collectives on AMD Infinity Fabric and HPC-Slingshot 11 GPU and node interconnects. This aims to understand the communication overhead for the underlying calls at the OMB microbenchmark level, using the same communication backend as employed by our training framework of choice, GPT-NeoX\cite{gpt-neox-library}.

\subsection{Contributions} 
Our contributions are as follows:
\begin{enumerate}
\item We combine empirical results with analytical models to study communication behavior for various parallelism schemes and sequence lengths.

\item We provide an in-depth understanding of the communication overheads associated with Data, Pipeline, and Tensor parallelism schemes commonly used in transformer models.

\item We present system-agnostic and system-specific measurements for each parallelism scheme, including collective communication types, data volumes, proportions, frequency, and message sizes.

\item We examine the impact of sequence length on communication volumes per collective pattern for Data-Parallel and Model-Parallel environments.

\item We conduct interconnect-specific evaluations, measuring latency and bandwidth for the particular collectives used by the studied LLM models. The analysis is conducted on AMD Infinity Fabric and HPE-Slingshot 11 GPU and node interconnects.
\end{enumerate}
\textbf{To the best of our knowledge, this is the first study to systematically characterize communication for distributed transformer models across multiple parallelism schemes and sequence lengths, providing detailed insights into collective communication types, data volumes, and distributions, and combining these results with the interconnect-specific collective communication benchmarking on the Frontier supercomputer.}
\vspace{-1.5ex}
\subsection{Paper Breakdown}
The rest of this paper is broken down as follows. Section \ref{sec:background} explains the background of LLMs and parallelism schemes used to train them and other DL models on HPC clusters. Section \ref{sec:performance-model} details the set of equations used to model communication volume for each parallelism scheme used in this paper. Sections \ref{sec:system-setup} and \ref{sec:results} break down our experimental results and how they relate to our performance model. Section \ref{sec:related} details related work in LLM characterization from its behavior to system-level performance. Section \ref{sec:conclusions} will conclude this paper and offer our suggestions and insights.
\vspace{-0.3ex}
\section{Background}
\label{sec:background}
\vspace{-1ex}
\subsection{Transformer Architecture}
\vspace{-.5ex}
The current trend in Natural Language Processing (NLP) favors transformer models \cite{vaswani2017attention} for their exceptional accuracy and computational efficiency. The original transformer architecture is designed for machine translation and contains two main components: an Encoder and a Decoder. Modern adaptations of transformers for language modeling utilize either the Encoder or Decoder depending on the specific task, such as BERT \cite{devlin2018bert} and GPT-2 \cite{radford2019language}.

A transformer layer is structured with a self-attention block followed by a two-layer multi-layer perceptron (MLP), composed of two GEMMs and a GeLU non-linearity (ReLU for the original version \cite{vaswani2017attention}). Each encoder or decoder block includes multiple such layers, each featuring multi-head attention, MLP, normalization, and residual connections.

We consider a single encoder or decoder with multiple transformer layers. Initially, input tokens are processed through a word embedding table and combined with positional embeddings, resulting in a 3-D tensor of size (sequence length × micro-batch size × hidden dimension) \cite{korthikanti2022reducing}. Each transformer layer processes this tensor through a self-attention block with multiple attention heads and a two-layer MLP that quadruples the hidden size and then reduces it back. The output size remains consistent across layers, and the final output is projected back to the vocabulary dimension for cross-entropy loss calculation. 

\subsection{Parallelism Techniques}
\vspace{-0.5ex}
Larger models are more sample-efficient given a fixed compute budget \cite{hoffmann2022training, kaplan2020scaling}, leading to a massive increase in model parameter count. Training billion/trillion-parameter transformer models is a memory-intensive task since it requires efficient distribution of multiple training parameters (model weights, optimizer states, gradients, and activations).

In \textbf{Data Parallelism} \cite{bennun2018demystifying}, a training mini-batch is divided among multiple workers and each worker maintains a full model replica. Data parallelism can achieve near-linear scaling in training data throughput by increasing the mini-batch size in proportion to the number of available workers. Typically, an Allreduce on all the workers is required to synchronize the gradients before updating the model weights on each local replica. Data Parallelism is communication-bound since the achievable bandwidth and latency of the Allreduce greatly affect iteration time given a worker's memory is consumed by the model and other training parameters. However, data parallelism requires that model size must fit in the limited GPU memory and additional optimizer and hyper-parameter tuning to ensure convergence with large global batch size \cite{you2020large}.

\textbf{Pipeline Parallelism} mainly focuses on distributing layers of models among GPU workers and executes these layers in a pipeline order. Since activation computation relies on dependencies between different layers, inevitable GPU idle times, known as pipeline bubbles are present in this paradigm, there have been various research efforts in reducing such bubbles \cite{huang2019gpipe, harlap2018pipedream}. In terms of communication, pipeline parallelism involves point-to-point GPU communication to pass along activations between layers.

\textbf{Tensor Parallelism} \cite{shoeybi2019megatron} aims at exploiting the inherent parallelism inside GEMM operations and distribute these computations along specific directions (rows, columns) and use synchronization among workers to gather the results, thus ensuring correctness. State-of-the-art implementations distribute the MLP blocks and Self-Attention blocks \cite{shoeybi2019megatron}. Results are collected and aggregated using Allreduce and Allgather. It is a common practice to limit tensor parallelism degree within a compute node since intra-node bandwidth is typically larger than inter-node bandwidth \cite{LLM_Scale_out_Char}. 

\begin{figure}[ht!]
    \centering
    \includegraphics[width=.7\columnwidth]{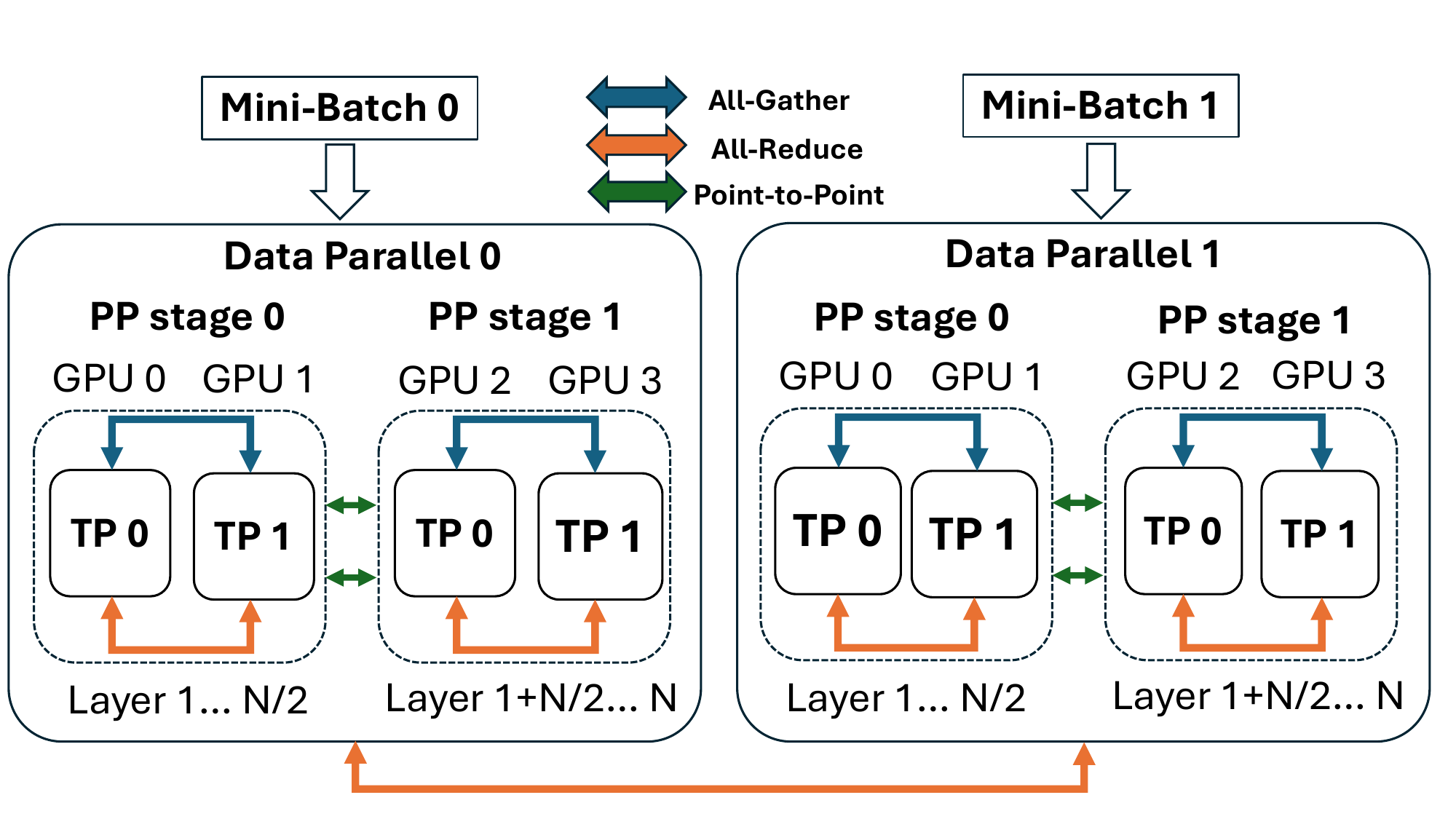}
\vspace{-2ex}
    \caption{An illustration of 3D parallelism with 2 Data-Parallel ranks, 2 Pipeline-Parallel stages and 2 Tensor-Parallel ranks. Each Pipeline-Parallel stage holds half of the total layers.}
    \label{fig:3d-topo}
\end{figure}

Figure \ref{fig:3d-topo} demonstrates \textbf{3D Parallelism}, which combines Data Parallelism, Pipeline Parallelism and Tensor Parallelism. This synergy has been a widely adopted approach to scale up transformer training to thousands of workers. It has the benefit of preventing global batch size from growing atrociously but requires effort to implement and prototype.
\vspace{-2ex}

\vspace{-2ex}

\subsection{Zero Redundancy Optimizer}
\label{sec:background-zero}
\vspace{-.7ex}
Data parallel training requires each rank to hold a copy of all model optimizer states, gradients, and parameters. \cite{ZeRO} Zero Redundancy Optimizer (ZeRO) reduces memory constraints by removing redundant information, and partitioning model data across data parallel ranks. ZeRO is divided into three stages, \textbf{ZeRO-1}, \textbf{ZeRO-2}, and \textbf{ZeRO-3}.
\begin{figure}[ht!]
    \centering
    \includegraphics[width=.7\columnwidth]{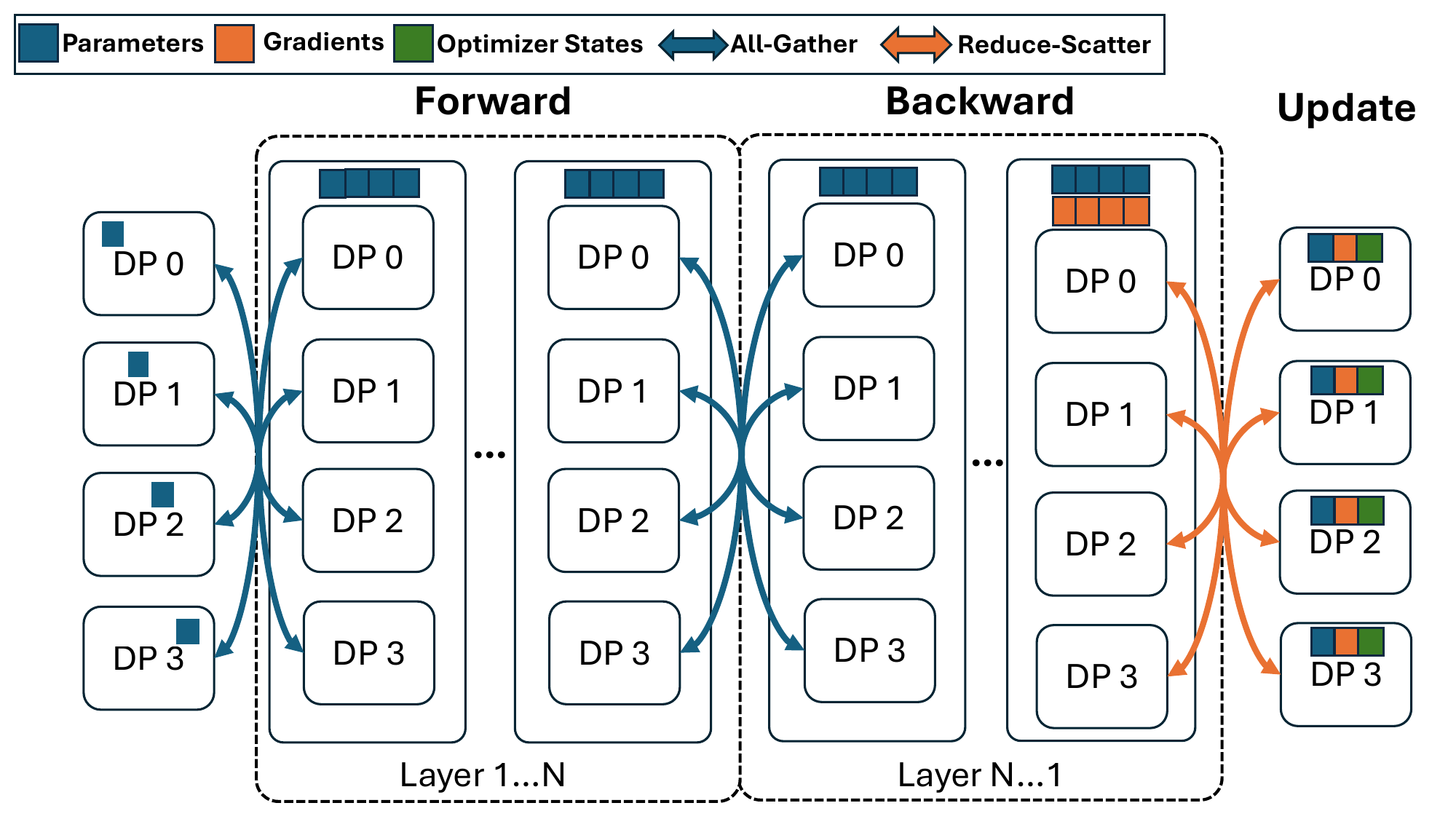}
\vspace{-1ex}
    \caption{An illustration of ZeRO-3 with 4 Data-Parallel ranks and $N$ layers. Between each layer, an Allgather is needed to collect the parameters from all the workers.}
    \label{fig:zero3-topo}
\end{figure}
Given a certain degree of data parallelism, each ZeRO stage partitions different training parameters. \textbf{ZeRO-1} partitions optimizer states across workers. Each worker only needs to store and update its partitions. At the end of each training step, an allgather is required to collect the fully updated model weights. \textbf{ZeRO-2} further partitions gradients and reduces them to only update the corresponding parameters. After gradient reduction, the memory can be released immediately, which will further alleviate memory pressure on a worker. Such a process requires Reduce-Scatter to distribute and reduce the gradients. ZeRO-1 and ZeRO-2 produce the same communication volume as standard data parallelism \cite{ZeRO}. \textbf{ZeRO-3} applies model parameter partitioning on top of optimizer states and gradients. However, stage 3 requires an extra allgather to collect parameters from all other processes as needed in forward and backward computation which typically incurs 1.5x communication volume compared to data parallelism baseline (Figure \ref{fig:zero3-topo}). 

\textbf{ZeRO++} applies various optimizations towards ZeRO-3, aiming at reducing communication volume and featuring a bandwidth-aware partitioning strategy. Specifically, ZeRO++ integrates blocked-based quantization kernels \cite{dettmers20228bit} into model weights and gradient communications to drastically reduce message size. It also keeps a secondary parameter partition within a compute node so that high-latency inter-node Allgather can be avoided due to low interconnect bandwidth \cite{wang2023zero}.
\section{Performance Model}
\label{sec:performance-model}
\vspace{-2ex}
\begin{table}[ht]
\resizebox{\columnwidth}{!}{%
\begin{tabular}{l|ll|l}\toprule
\textit{a} & Number of attention heads    & \textit{s} & Sequence length      \\
\textit{b} & Microbatch size              & \textit{t} & Tensor-parallel size \\
\textit{h} & Hidden dimension size        & \textit{V} & Vocabulary size     \\
\textit{L} & Number of transformer layers & \textit{p} & Pipeline-parallel size \\
\textit{d} & Number of training devices
\end{tabular}%
}
\caption{Variable names.}
\label{tab:varnames}
\end{table}

This section breaks down each component that makes up our performance model.

\subsection{Data Parallelism and ZeRO}

To calculate the total parameters in a transformer, we have the embedding and unembedding blocks of size $V \times h$ each. If embedding and unembedding parameters are tied (i.e. shared), this leads to a total of $V \times h$ parameters from embeddings. Since all configurations in this paper use untied embeddings, we have $2V \times h$ embedding parameters. We also have the position embeddings of size $sh$. The attention matrices are four separate matrices of dimension $h\times h$, leading to $4h^2$ attention parameters per layer. Multilayer perceptron (MLP) blocks for our models are composed of two fully-connected linear projections of size $h \times xh$ and $xh \times h$, where $x$ is the \textit{expansion factor}. For GPT-NeoX model architectures, the conventional projection factor is $4$~\cite{LLM-Arch}, so we have $2xh^2 = 8h^2$ MLP parameters per layer. We then have a layernorm each layer with both gains and biases on each of the $Q, K, V$ and the first MLP linear projection, leading to $8h$ layernorm parameters per layer. Finally, we add the final layernorm of size $2h$ to get a total number of parameters in Equation \ref{eq:param_count} below.  

\vspace{-1ex}
\begin{equation}
    param\_count = 2Vh + sh + L(12h^2 + 8h) + 2h
    \label{eq:param_count}
\end{equation}

Considering a message size of $m$, the communication volume for the Allreduce collective is $2\times m (\frac{d-1}{d})$. The communication volume for Allgather, Reduce\_scatter, and Reduce is simply $m (\frac{d-1}{d})$.

The communication volume per iteration for distributed data parallelism (DDP) just comes from the gradient Allreduce, which gives the total volume per iteration given in Equation \ref{eq:DDP-ZeRo-1-2} below. ZeRO-1 and ZeRO-2 simply replace this Allreduce call with separate Reduce\_scatter and Allgather calls~\cite{ZeRO}, so they have the same communication volume as DDP. Therefore, the communication volume (in units of parameters) from DP (Allreduce), ZeRO-1, and ZeRO-2 (Allgather/Reduce\_scatter) is given by:
\vspace{-1.5ex}
\begin{equation}
    2*param\_count * (\frac{d-1}{d})
    \label{eq:DDP-ZeRo-1-2}
\end{equation}

The communication volume for ZeRO-3 is 50\% higher due to an extra Allgather of parameters, which is necessary before the forward pass because parameters are now also sharded across ranks (See \ref{sec:background-zero} and \cite{ZeRO}). Therefore, the ZeRO-3 communication volume (in units of parameters) is given by:
\vspace{-1.5ex}
\begin{equation}
    3*param\_count * (\frac{d-1}{d})
    \label{eq:ZeRO-3}
\end{equation}

\subsection{Model Parallelism}

The communication volume for pipeline parallelism comes from the point-to-point communication of forward activations and backward gradients. The send or receive between two pipeline stages is of size $bsh$, therefore the aggregate communication volume across all stages in a single training iteration is given in Equation \ref{eq:Pipeline_par} below (in units of parameters and where $d$ is the number of devices, or GPUs, used in training). Notably, the first stage doesn't have to receive activations and the last GPU doesn't have to send activations (and vice-versa with gradients), so we multiply by $p-1$ instead of $p$.

\vspace{-1.5ex}
\begin{equation}
    2bsh \times (p-1)
    \label{eq:Pipeline_par}
\end{equation}

The communication volume per iteration for tensor parallelism comes from 6 Allreduce operations per layer (2 in the forward pass, 2 for activation recomputation, 2 in the backward pass). Further, an additional Allreduce operation is performed at the embedding. Each Allreduce incurs a volume of $2m$, leading to a total of $(12L + 2)$ volume for messages of size $bsh$. Since these Allreduce operations are across $t$ ranks, they're multiplied by a factor of $\frac{t-1}{t}$.

\vspace{-1ex}
\begin{equation}
    (12L+2) * bsh * (\frac{t-1}{t})
    \label{eq:Tensor_Par}
\end{equation}

For 3D parallelism, one simply updates the tensor parallelism equation to be $L \rightarrow L/p$. This implies that the total communication volume here is additive. 





\section{System Setup}
\label{sec:system-setup}
\begin{table}[]
    \centering
    \begin{tabular}{c|c}
    \toprule
        CPU &  AMD Epyc 7713 ``Trento" 64 core 2 GHz \\
        GPU &  4 x AMD MI-250X \\
        Interconnect & HPE Slingshot 11 (4 NICS/Node)\\
        ROCm Version Used & 5.6.0 \\
        CPU/GPU-Interconnect & AMD Infinity Fabric\\
        PyTorch Version Used & 2.1.2 \\
        DeepSpeed Version Used & 0.14 \\
        GPT-NeoX Version Used & commit 4bc667031d8 \\
        Dataset Used & enwik8 \\
    \bottomrule
    \end{tabular}
\vspace{-1ex}
    \caption{Experiment Setup Specifications}
    \label{Tabl:Frontier}
\end{table}

\begin{figure}[ht!]
    \centering
    \includegraphics[width=.76\columnwidth]{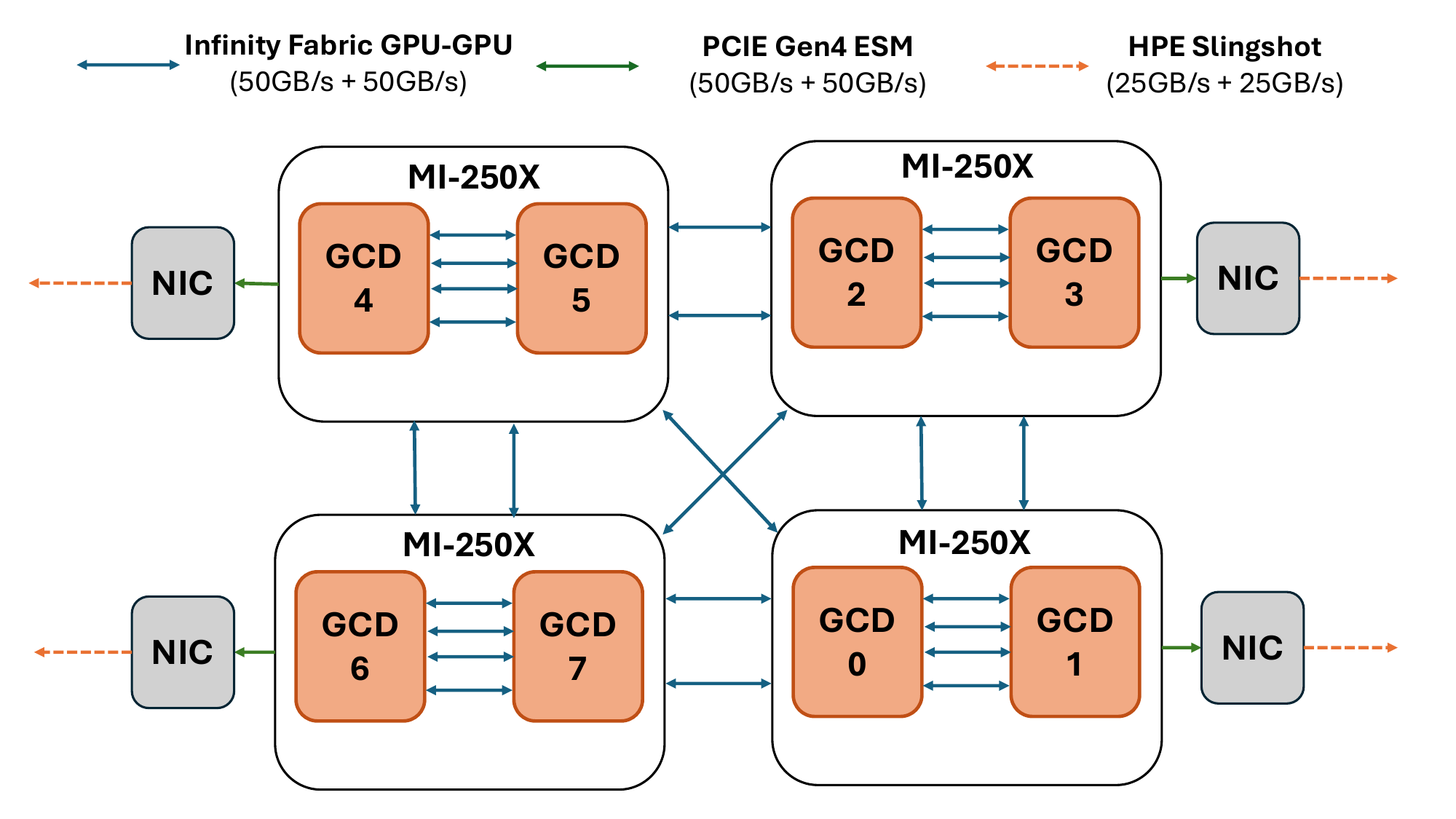}
\vspace{-1ex}
    \caption{Topology of a compute node on Frontier}
    \label{fig:frontier-topo}
\end{figure}

This section explains the experiments run, and insights gained from our results. All experiments were run on the OLCF Frontier supercomputer. See Table \ref{Tabl:Frontier} for more information on hardware and software specifics. For details on Frontier compute node topology, please refer to Figure \ref{fig:frontier-topo}. Regarding the use of Microsoft's DeepSpeed: we would like to note that communication/compute overlap is not possible when logging is turned on, which allowed us to obtain communication results featured in Section \ref{sec:results} with the following profiling numbers.

To facilitate easier training of the models involved, we utilize EleutherAI's ``GPT-NeoX" framework\cite{gpt-neox-library} and its configuration files for 19-million, 125-million, 1.3-billion, and 13-billion parameter models. The ``enwik8" dataset used features a vocabulary size of 50304 after padding to help with reducing performance runtime anomalies.

\section{Performance Characterization}
\label{sec:results}

\subsection{Data-Parallel Experiments (DDP, ZeRO-1/2/3)}
\label{sec:DDP-Zero}
Here, we explore the communication behavior of different Data-Parallel schemes such as pure data parallelism or different levels of DeepSpeed's ZeRO\cite{ZeRO}. Per the cost models referenced in Section~\ref{sec:performance-model}, DDP and ZeRO-1 and 2 should approximately achieve a volume proportional to twice the parameter count, and ZeRO-3 should achieve a communication volume equal to three times that of the parameter count. 


\begin{figure*}[ht!]
    \centering
    \begin{subfigure}[t]{0.2\textwidth}
    \includegraphics[width=\linewidth]{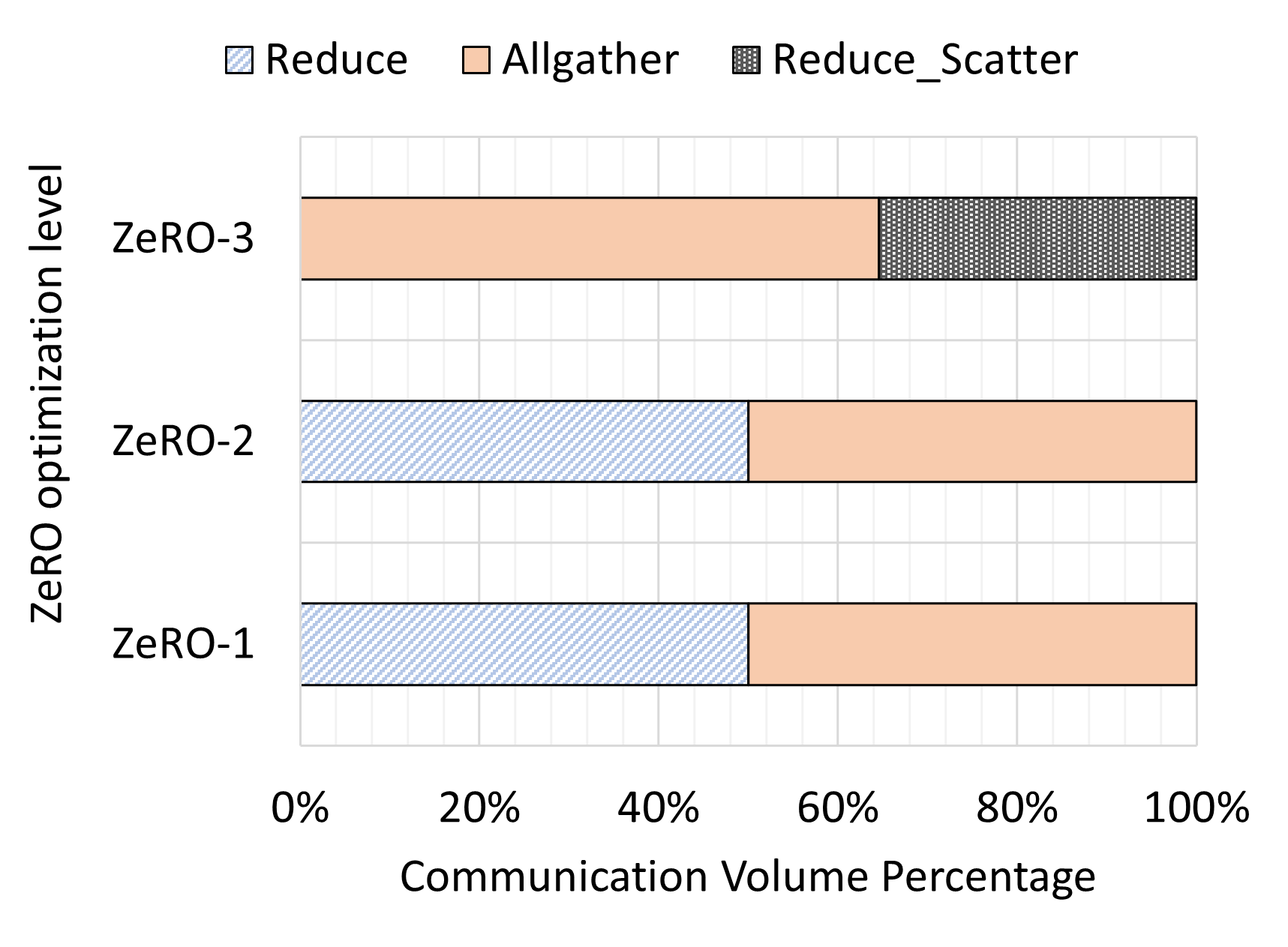}
    \caption{19M}
  
\end{subfigure}
\begin{subfigure}[t]{0.2\textwidth}
    \includegraphics[width=\linewidth]{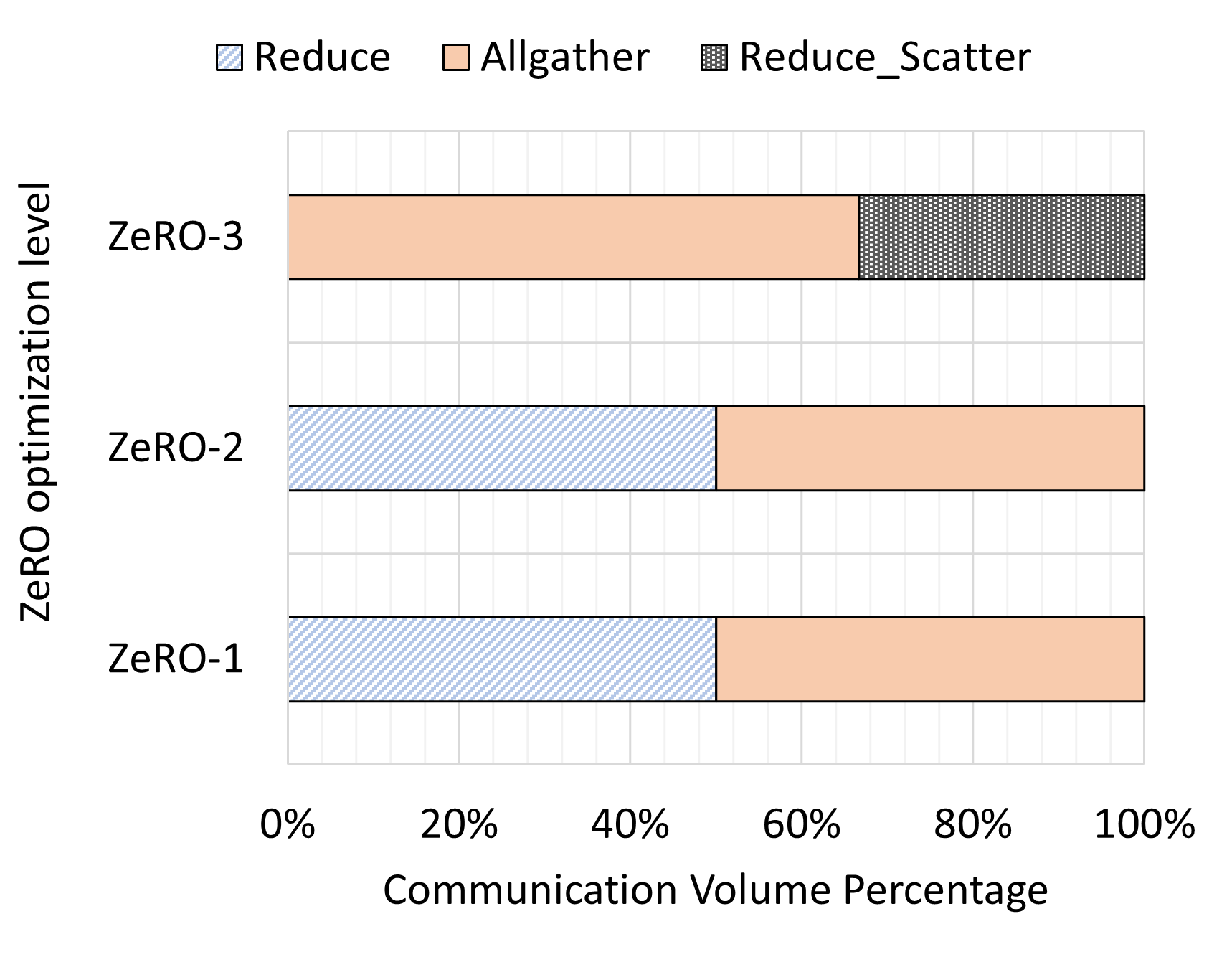}
    \caption{125M}
\end{subfigure}
\begin{subfigure}[t]{0.2\textwidth}
    \includegraphics[width=\linewidth]{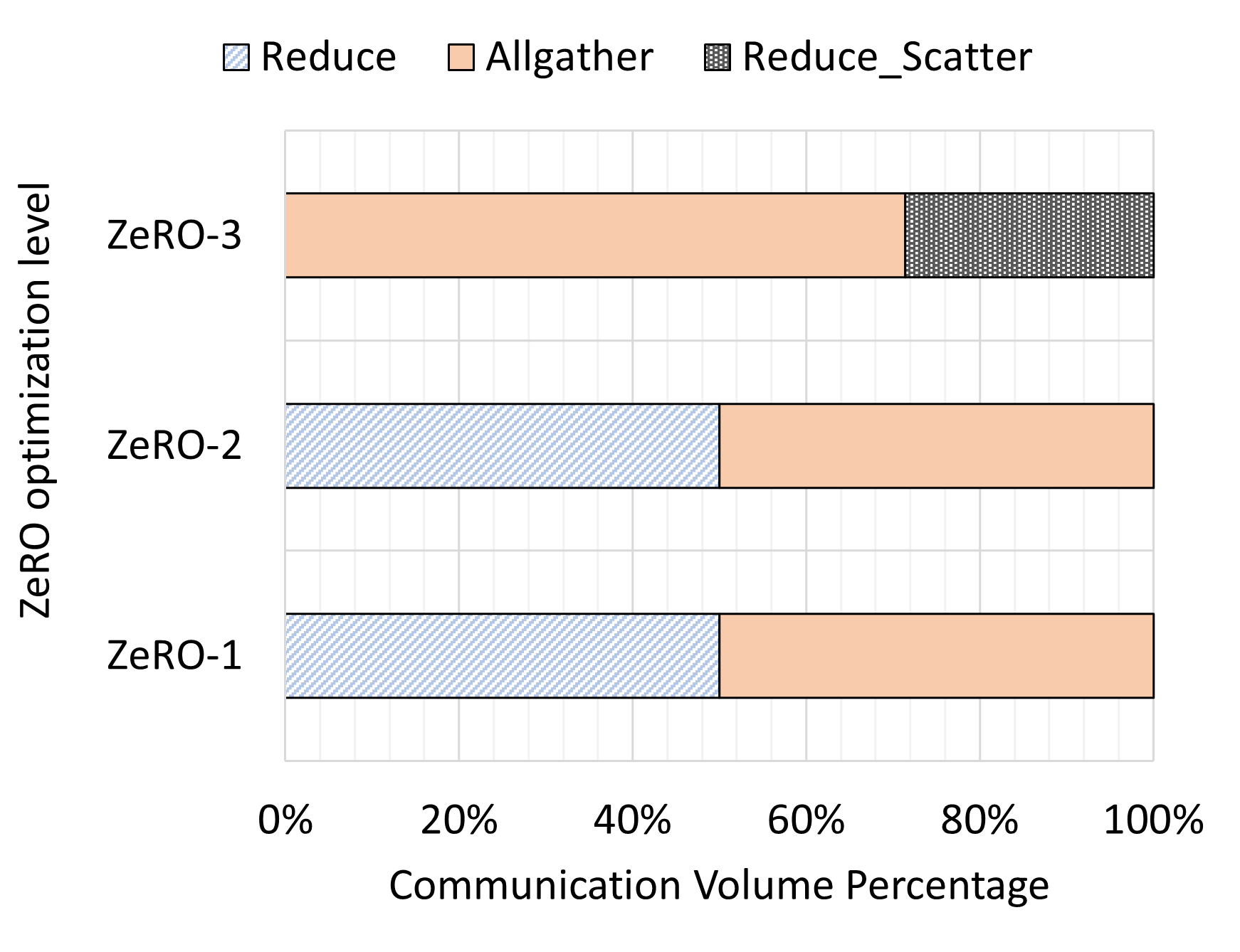}
    \caption{1.3B}
\end{subfigure}
\begin{subfigure}[t]{0.2\textwidth}
    
    \includegraphics[width=\linewidth]{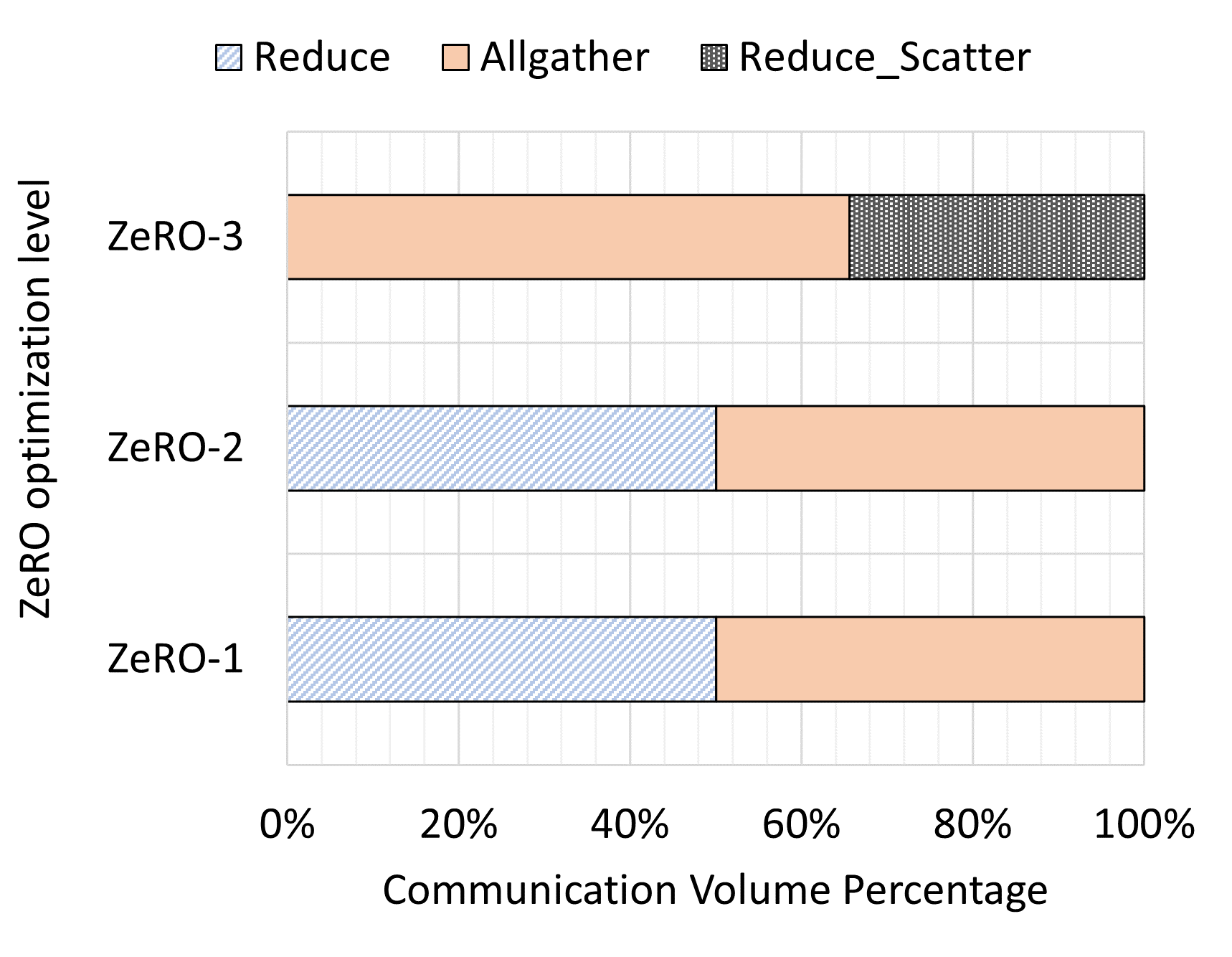}
    \caption{13B}
\end{subfigure}
\vspace{-1ex}
\caption{ZeRO-1/2/3 communication percentage breakdown for models of size 19M, 125M, 1.3B, and 13B.}
\label{fig:DP_Percentage}
    
\end{figure*}

\subsubsection{Breakdown of Communication Volume: ZeRO differences}

\begin{figure*}[ht!]
    \centering
    \begin{subfigure}[t]{0.2\textwidth}
    
    \includegraphics[width=\linewidth]{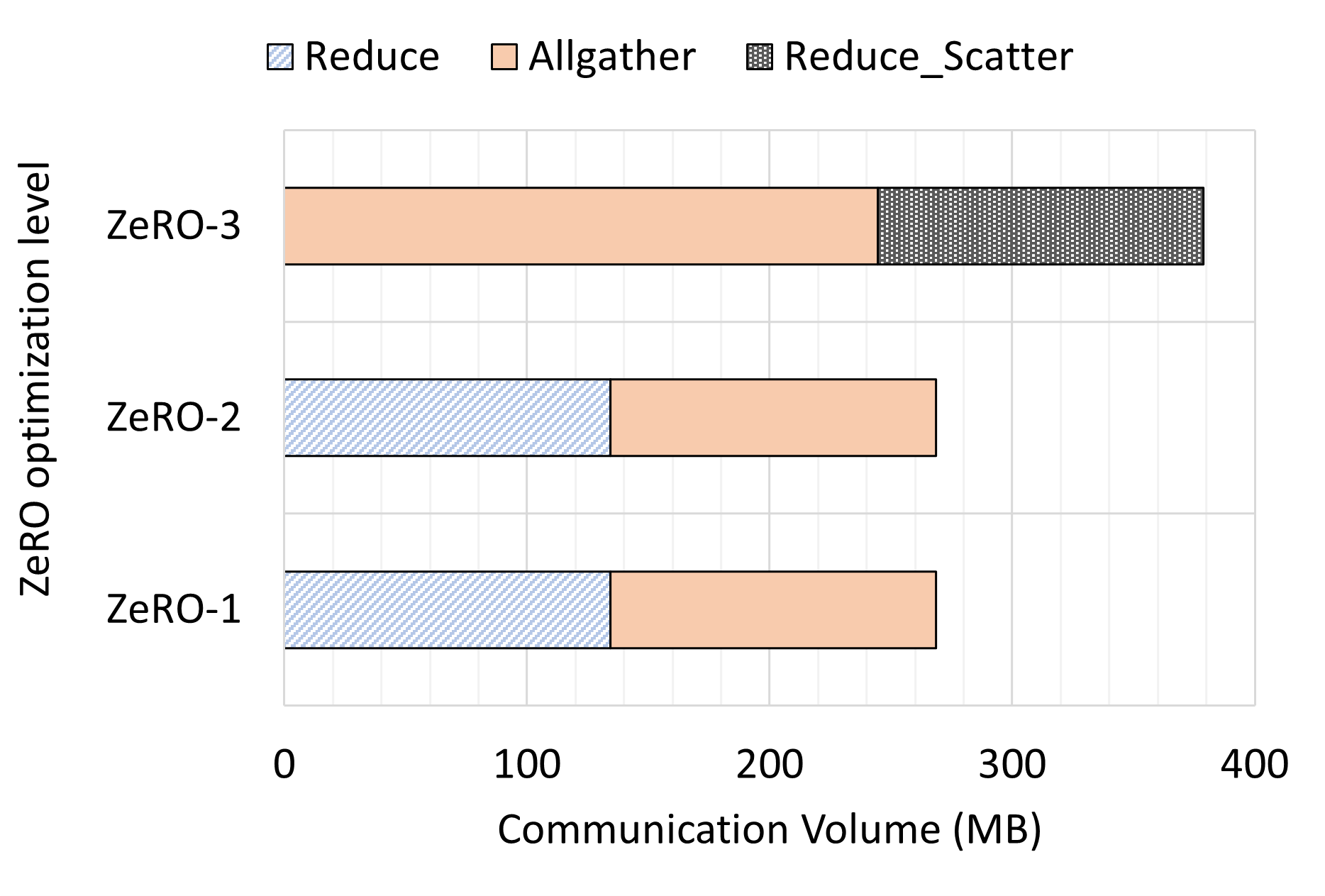}
    \caption{19M}
\end{subfigure}
\begin{subfigure}[t]{0.2\textwidth}
    
    \includegraphics[width=\linewidth]{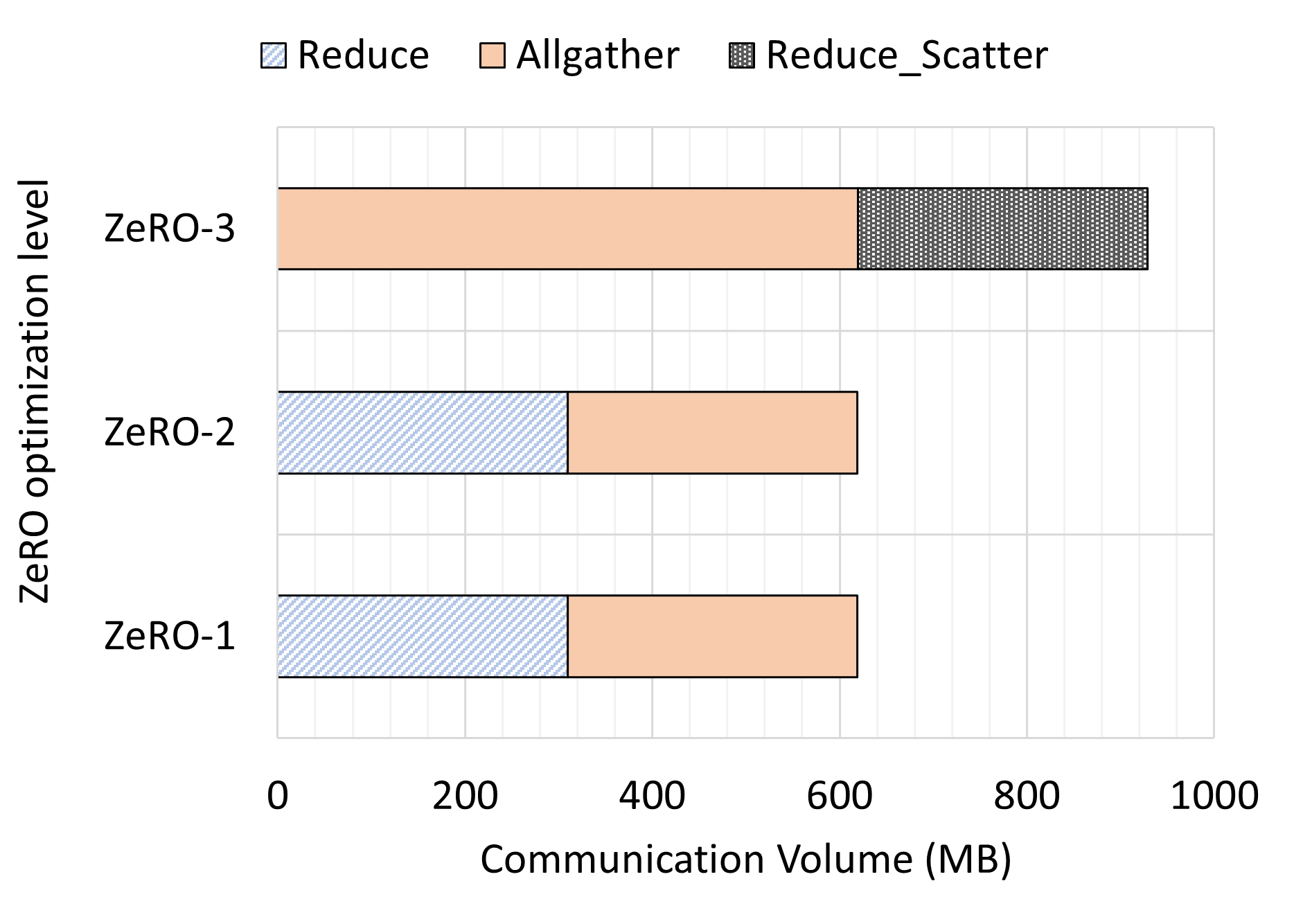}
    \caption{125M}
\end{subfigure}
\begin{subfigure}[t]{0.2\textwidth}
    
    \includegraphics[width=\linewidth]{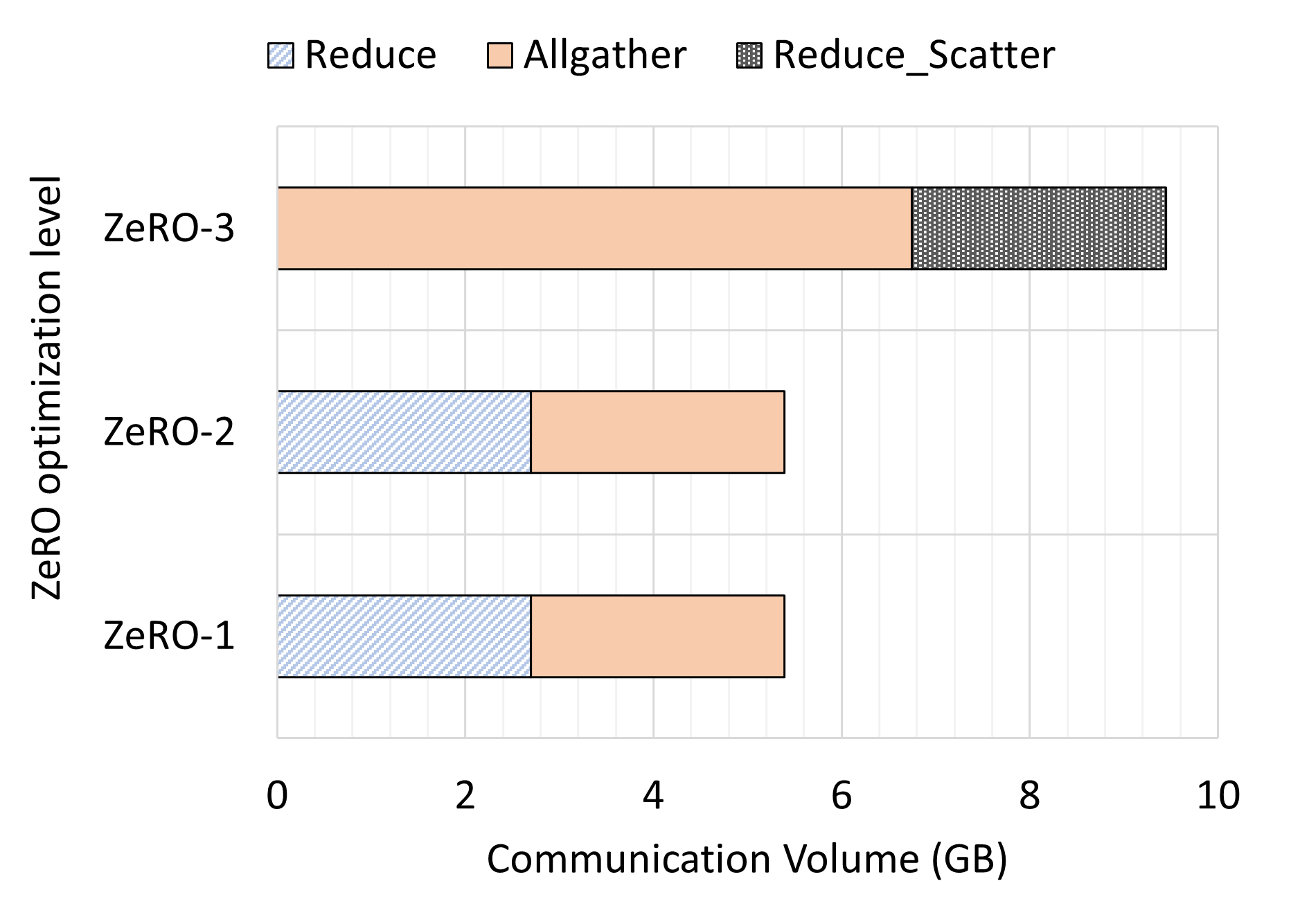}
    \caption{1.3B}
\end{subfigure}
\begin{subfigure}[t]{0.2\textwidth}
    
    \includegraphics[width=\linewidth]{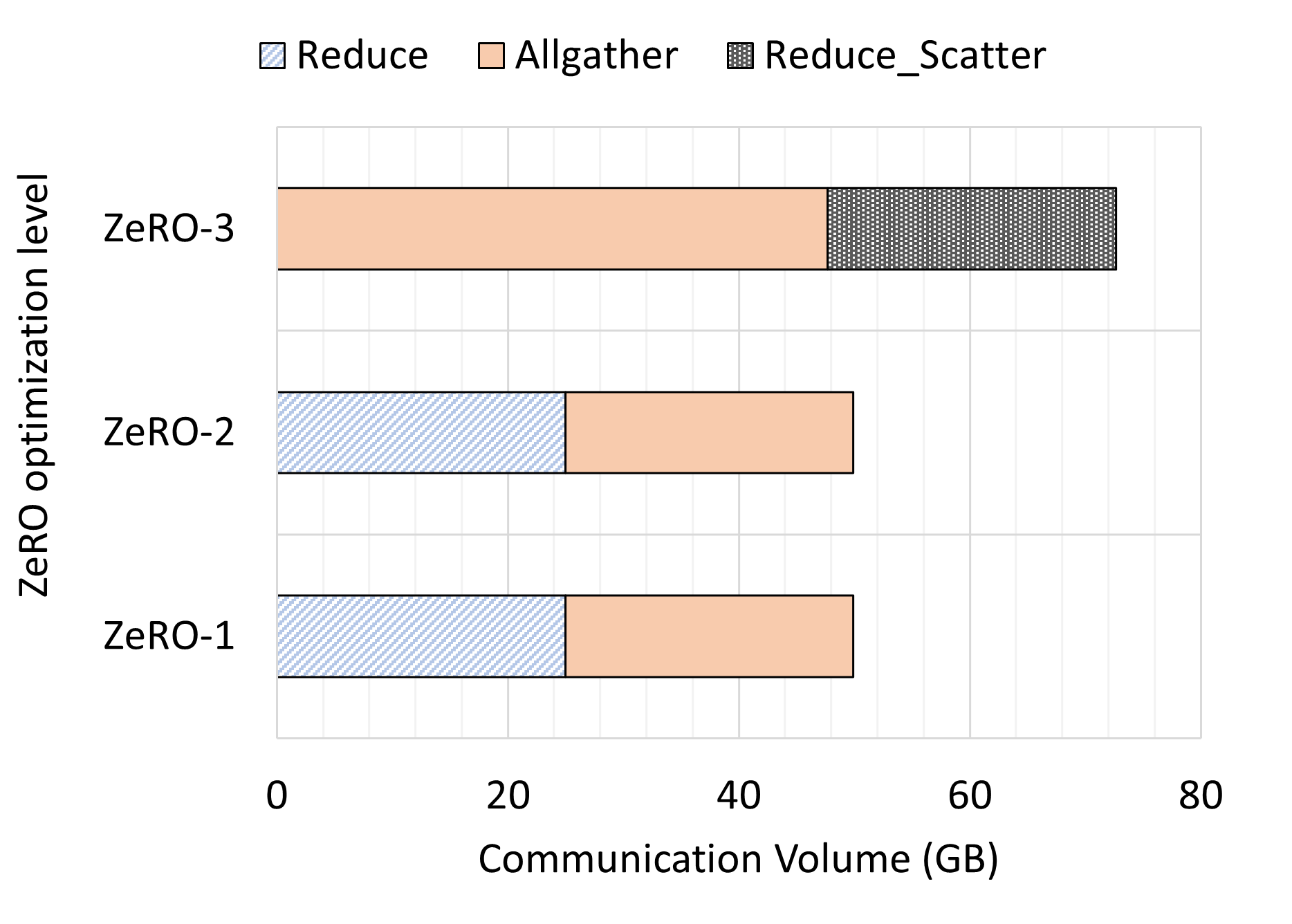}
    \caption{13B}
\end{subfigure}
\vspace{-1ex}
\caption{ZeRO-1/2/3 total communication volume for models of size 19M, 125M, 1.3B, and 13B.}
\label{fig:DP_Breakdown}
\end{figure*}

Figure \ref{fig:DP_Breakdown} shows communication breakdowns of each selected model size using one of ZeRO-1/2/3 (run on one node for all models except the 13B-parameter model due to memory errors. The models, as shown later still accurately hold up regardless of scale for a given model size). We want to note that Broadcast is included as a notion to the start-of-training parameter broadcast/distribution required, as this still incurs a level of overhead during initialization. Allreduce is still a significant portion of the communication in ZeRO-1/2 thanks to the fact that, aside from the 13B-parameter model, all other models can easily fit onto one of Frontier's MI250X GPUs with DDP. We would also like to note the general trend of decreasing broadcast impact as the model size increases, and this is also shown in Figure \ref{fig:DP_Percentage}, where each breakdown is them modeled as a percentage of the total communication volume.
\vspace{-0.4ex}
\subsubsection{Breakdown of Message Sizes and Frequency}
\begin{figure*}[ht!]
    \centering
    \begin{subfigure}[t]{0.33\textwidth}
    \includegraphics[width=\linewidth]{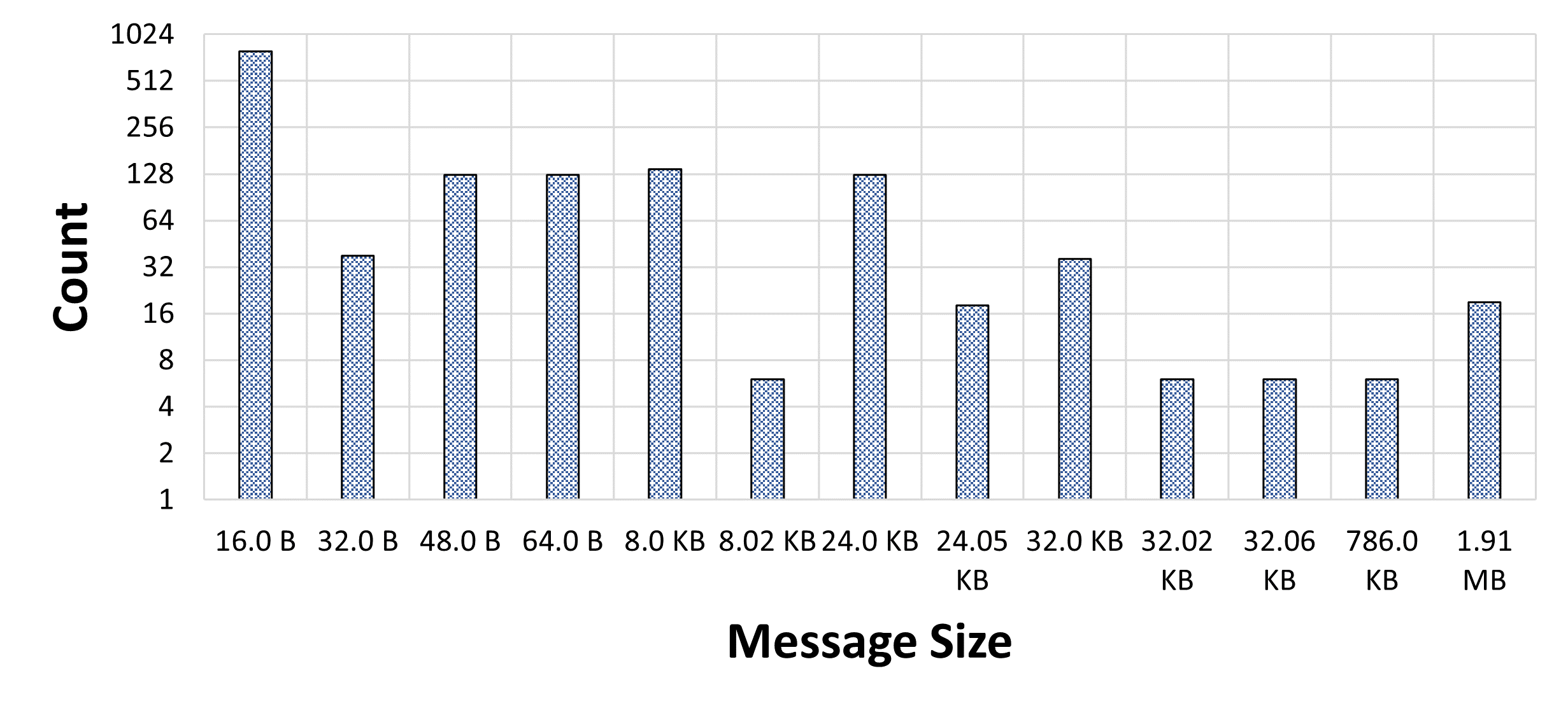}
    \caption{Allgather-message frequency breakdown, 19M-parameter model}
  
\end{subfigure}
\begin{subfigure}[t]{0.31\textwidth}
    \includegraphics[width=\linewidth]{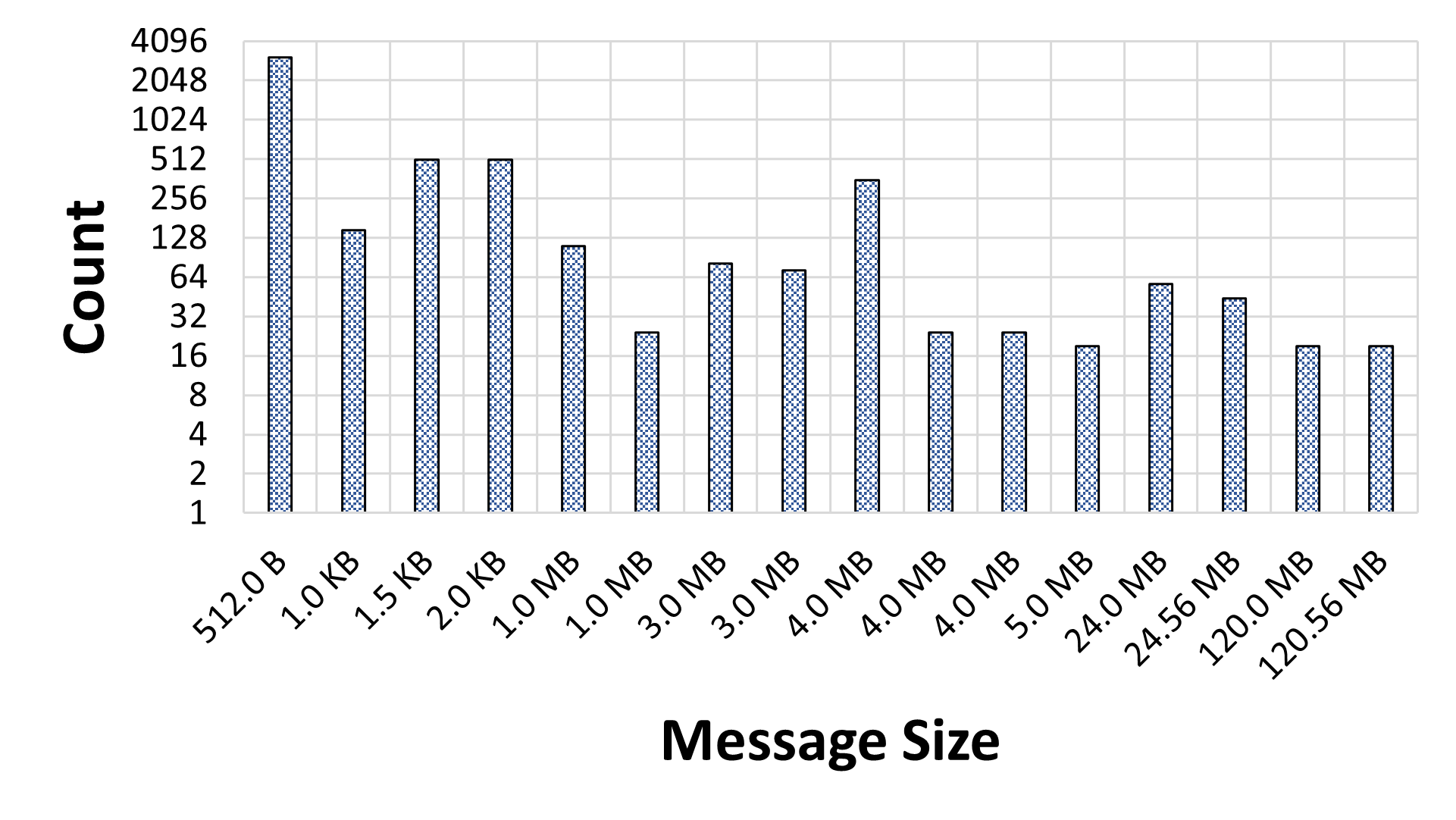}
        \caption{Allgather-message frequency breakdown, 1.3B-parameter model}

\end{subfigure}
\begin{subfigure}[t]{0.34\textwidth}
    \includegraphics[width=\linewidth]{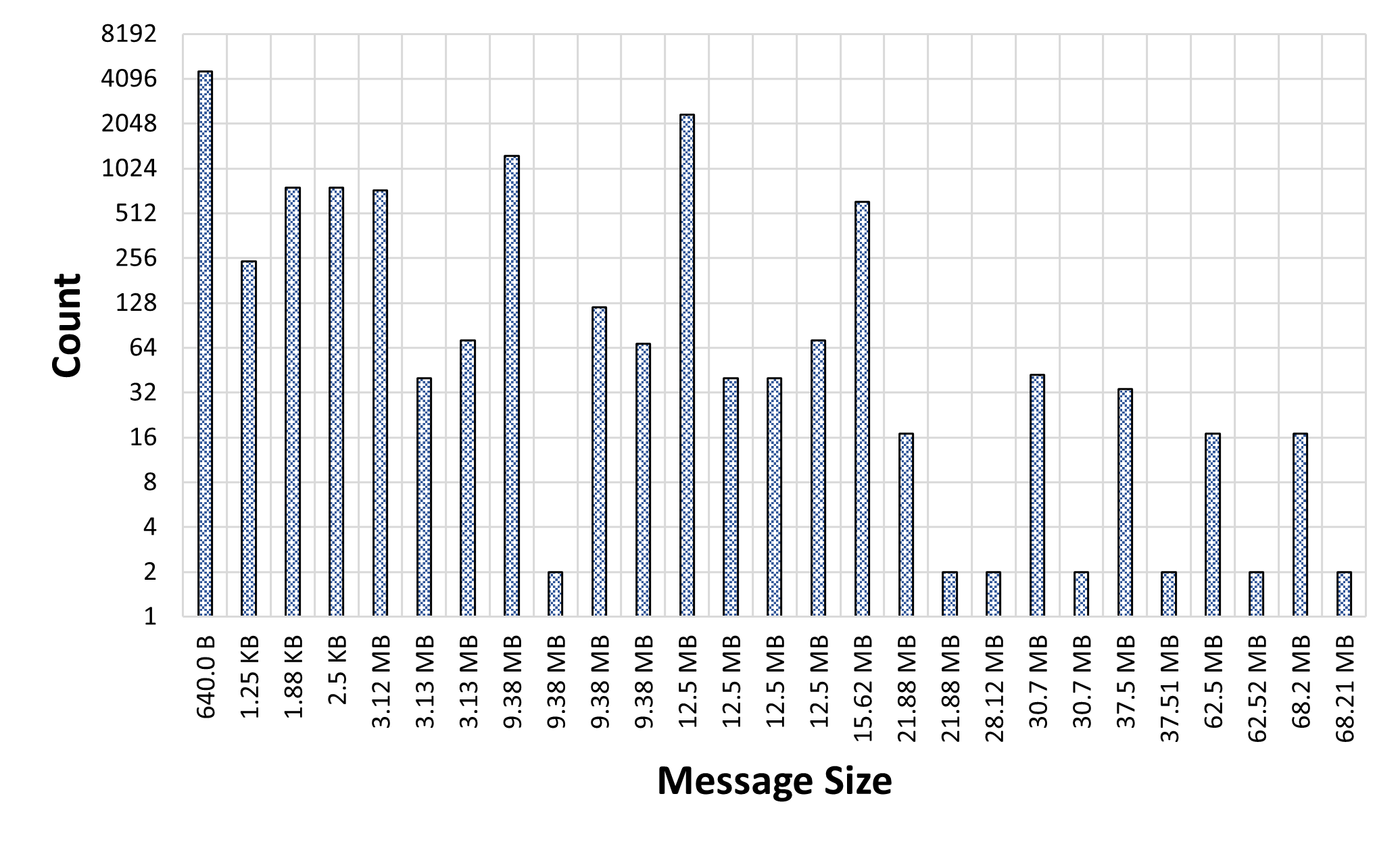}
        \caption{Allgather-message frequency breakdown, 13B-parameter model}
\end{subfigure}
\vspace{-1ex}
\caption{Message size breakdown for Allgather in three different model sizes utilizing ZeRO-3}
\label{fig:Allgather_freq}
\end{figure*}

As the model size increases, more message sizes for each communication call will be utilized, and to varying frequency levels. Figure \ref{fig:Allgather_freq} showcases 2-Node, 8 GCDs/Node experiments for 19-million, 1.3-billion, and 13-billion parameter models while using ZeRO-3. More verbose logging from DeepSpeed shows how message sizes get grouped into different categories for different functions; in the case of the 1.3-billion parameter model, many of the smaller messages (on the order of kilobytes) are used for parameter exchange among each process. Larger messages --- from 10s to 100s of megabytes --- are used for gradient aggregation (instead of an Allreduce as done in pure data parallelism). \textbf{The main takeaway: Even though DL models such as LLMs operate using massive message sizes, optimizations at smaller message sizes should be treated as equally important.}

\subsubsection{Comparison to Performance Model}
\begin{figure*}[ht!]
    \centering
    \begin{subfigure}[t]{0.24\textwidth}
    
    \includegraphics[width=\linewidth]{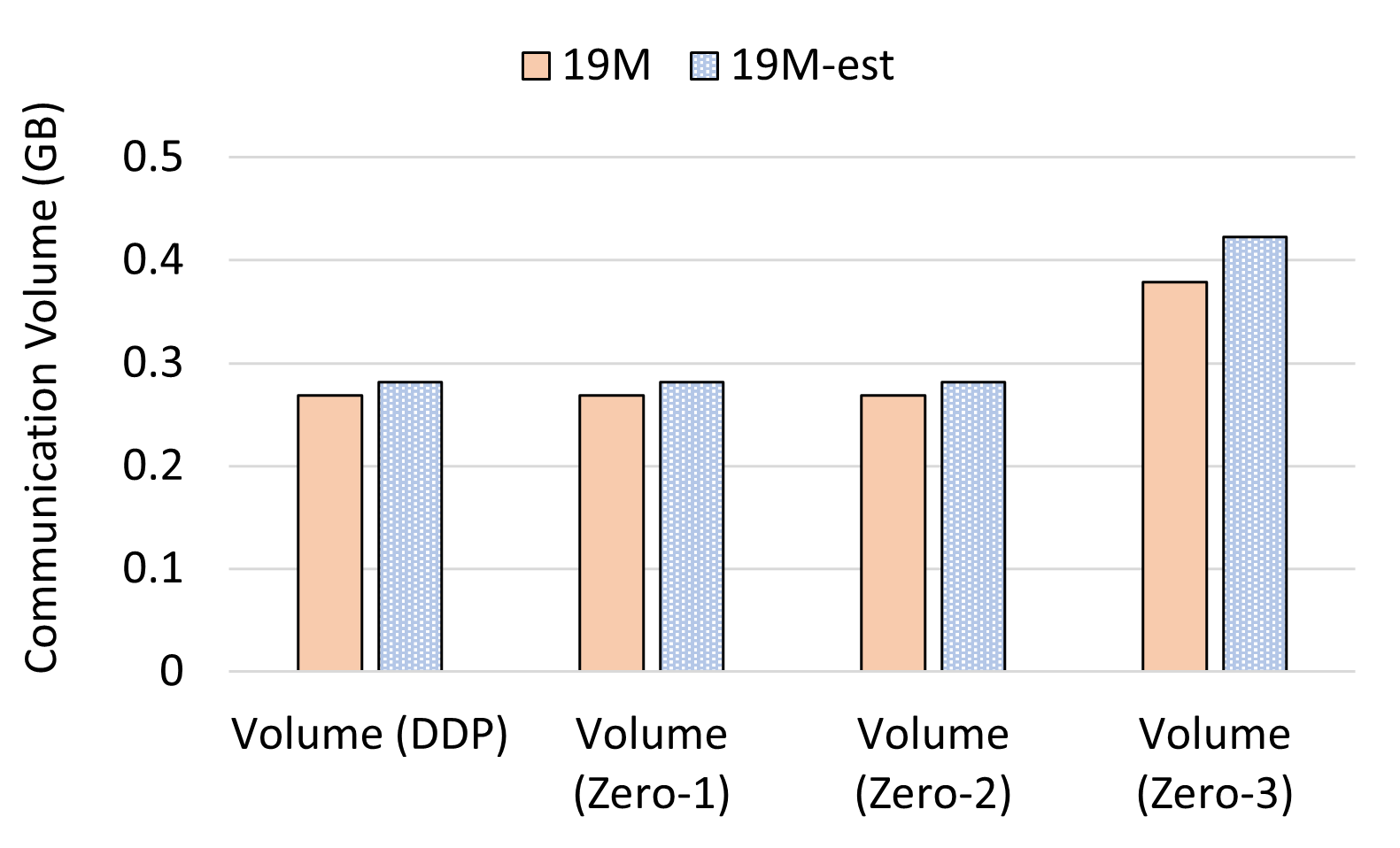}
        \vspace{-1ex}
        \caption{19M}
        \label{fig:19m-zero123}
    \end{subfigure}    
    \begin{subfigure}[t]{0.24\textwidth}
        \includegraphics[width=\linewidth]{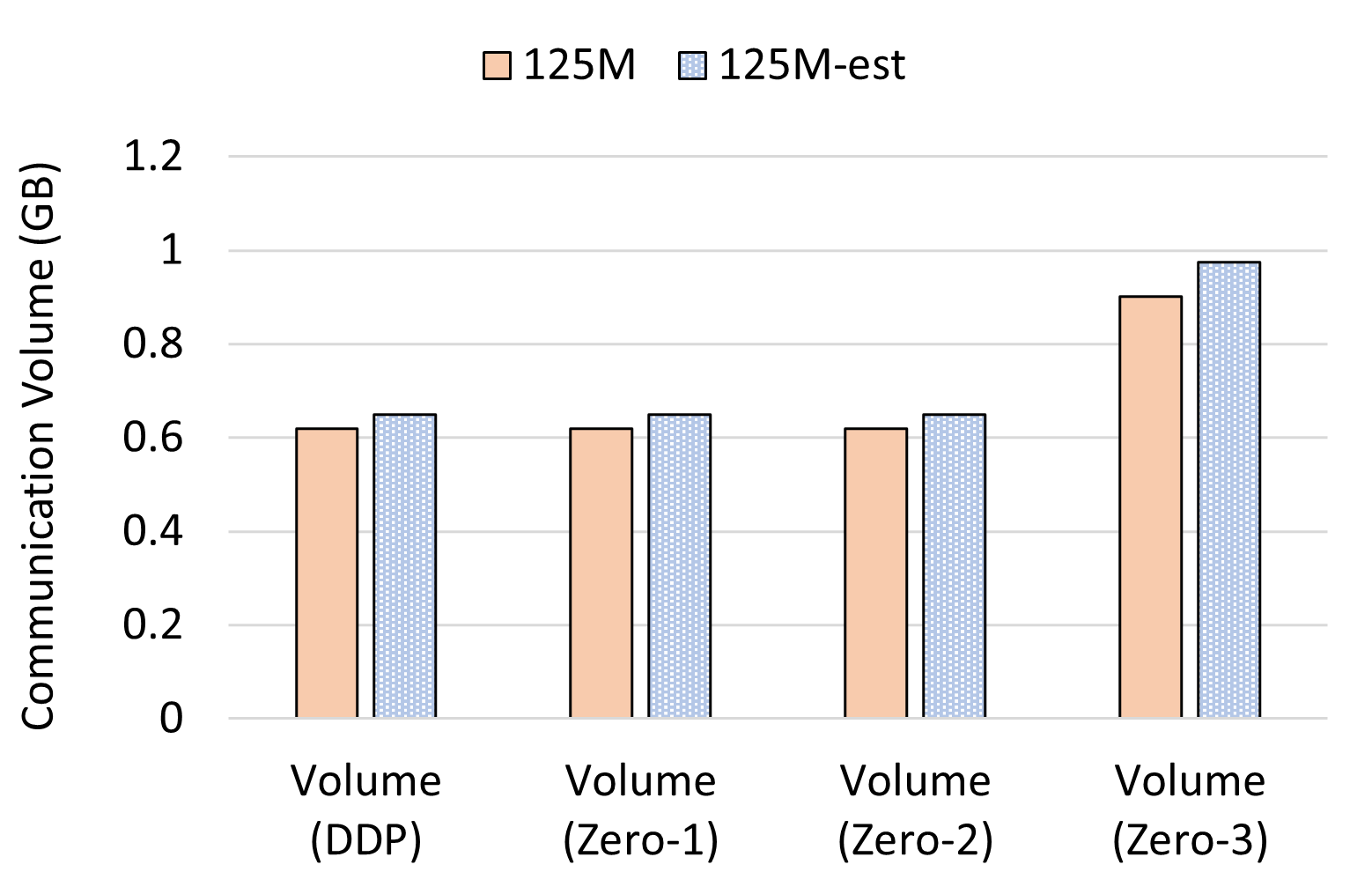}
        \vspace{-1ex}
        \caption{125M}
        \label{fig:125m-zero123}
    \end{subfigure}
    \begin{subfigure}[t]{0.24\textwidth}  
        \includegraphics[width=\linewidth]{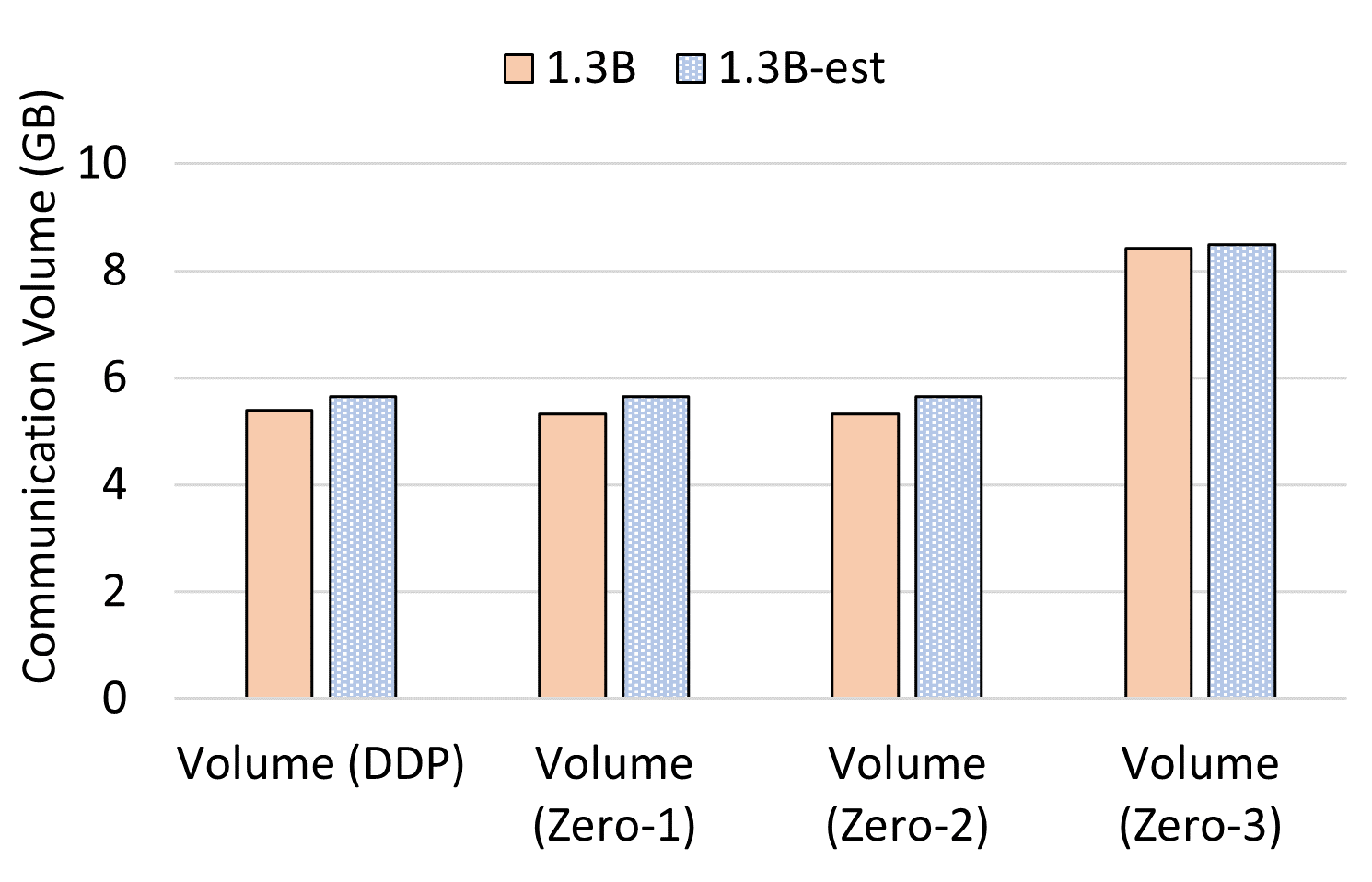}
        \vspace{-1ex}
        \caption{1.3B}
        \label{fig:1p3B-zero123}
    \end{subfigure} 
    \begin{subfigure}[t]{0.24\textwidth}  
        \includegraphics[width=\linewidth]{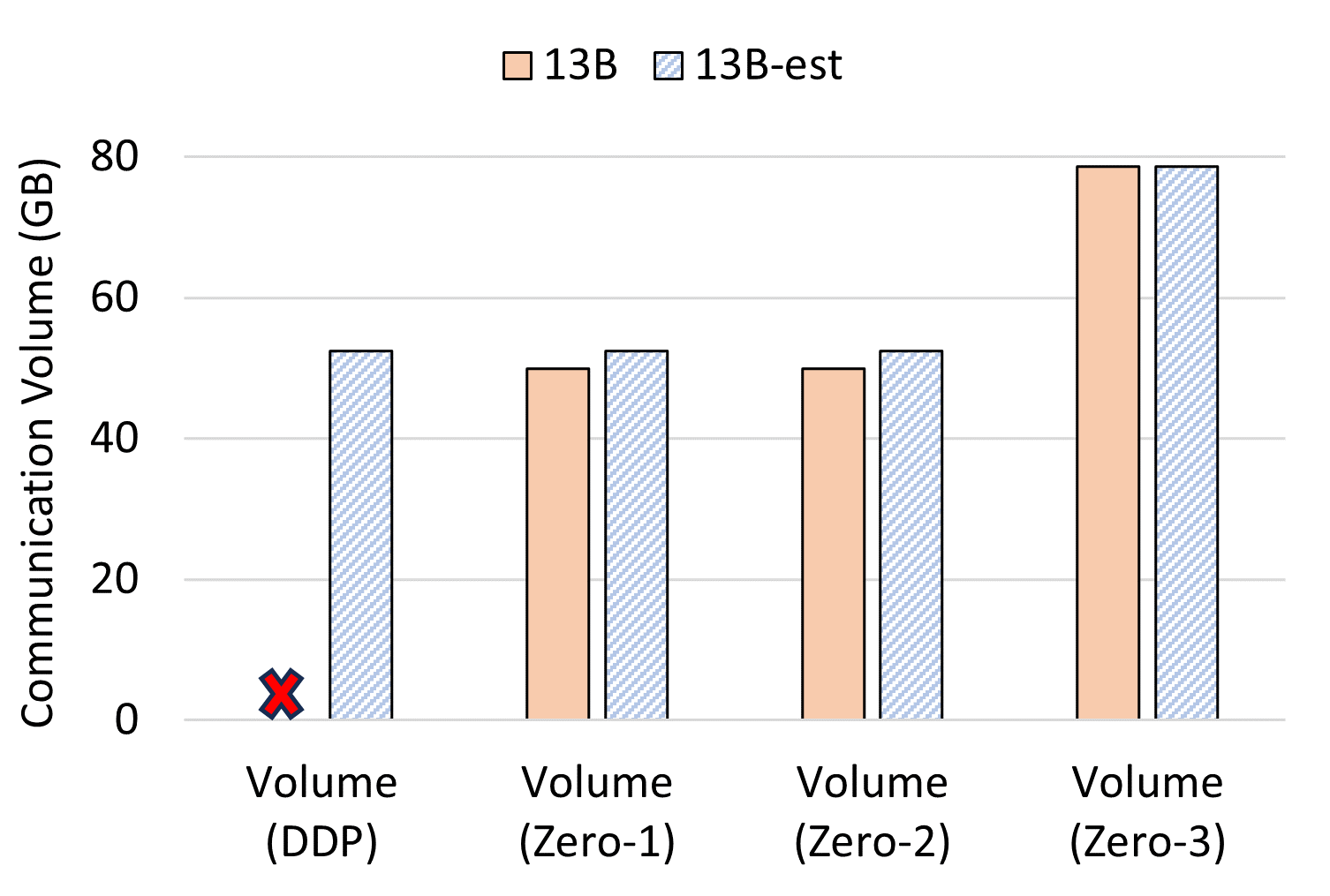}
        \vspace{-1ex}
        \caption{13B}
        \label{fig:13B-zero123}
    \end{subfigure}
    \caption{Communication volume for ZeRO-1/2/3 across model sizes 19M, 125M, 1.3B, and 13B}
    \label{fig:zero-volume-123}
\end{figure*}

Figure \ref{fig:zero-volume-123} shows how the 19M, 125M, 1.3B, and 13B-parameter models match up to the predicted communication volumes based on the Data-Parallel and ZeRO-based formulas from Section \ref{sec:performance-model}. In general, our prediction aligns well with the communication volume observed across all model sizes and all parallelism schemes (DDP, ZeRO-1/2/3). Note that we are able to predict 13B communication volume under a Distributed Data-Parallel scenario but training parameters will exceed worker memory in action, causing an OOM error.

\subsection{Model Parallelism Communication Volume Analysis (Tensor and Pipeline)}
\label{subsec:ModelPar}
This section explores the differing communication behaviors for tensor/pipeline parallelism and a combination of them in parallel (model parallelism).

\subsubsection{Breakdown of Communication Volume}
\begin{figure*}[ht!]
    \centering
    \begin{subfigure}[t]{0.35\textwidth}
    
    \includegraphics[width=\linewidth]{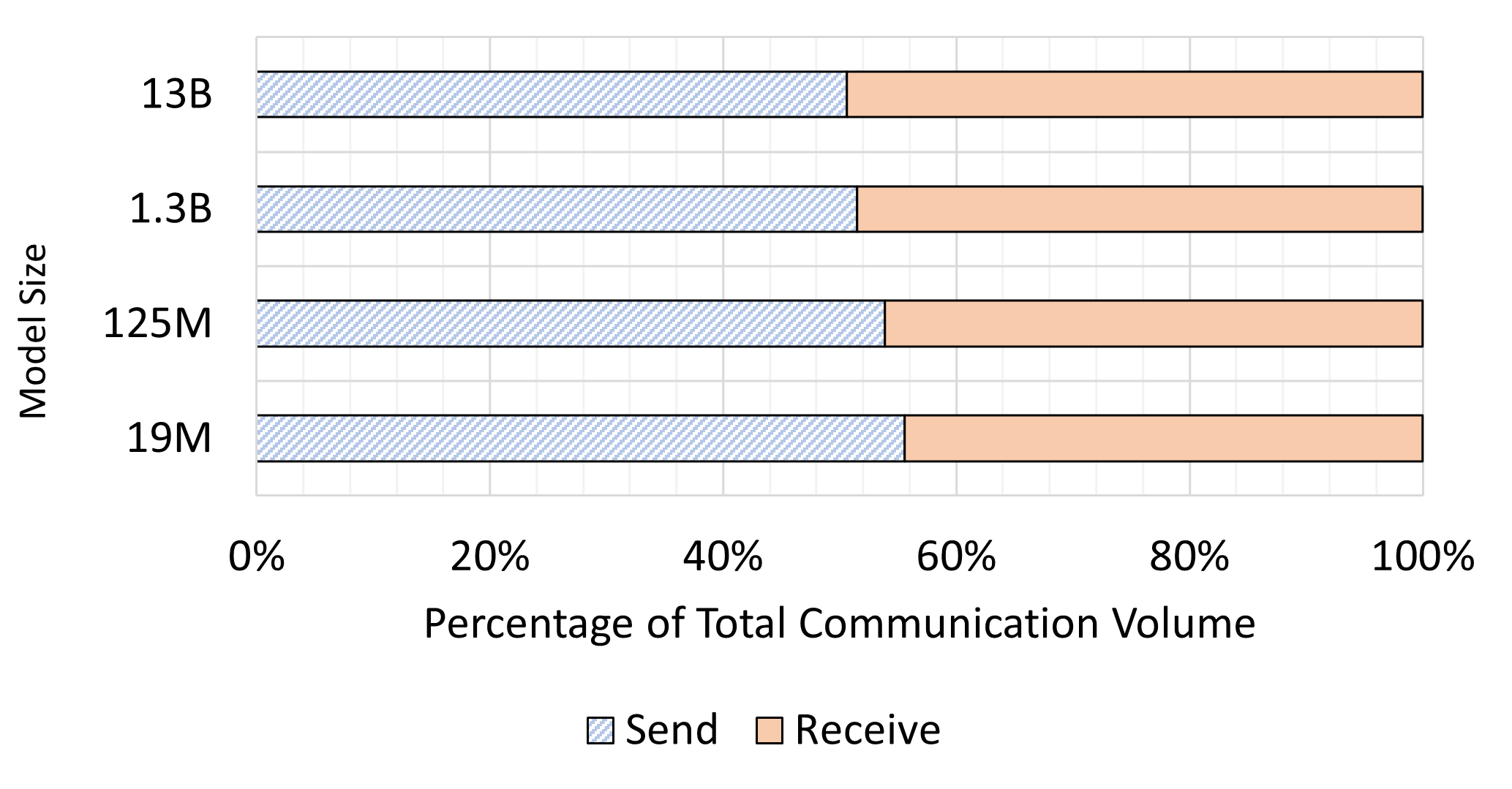}
    \caption{Pipeline Parallelism}
\end{subfigure}
\begin{subfigure}[t]{0.35\textwidth}
    
    \includegraphics[width=\linewidth]{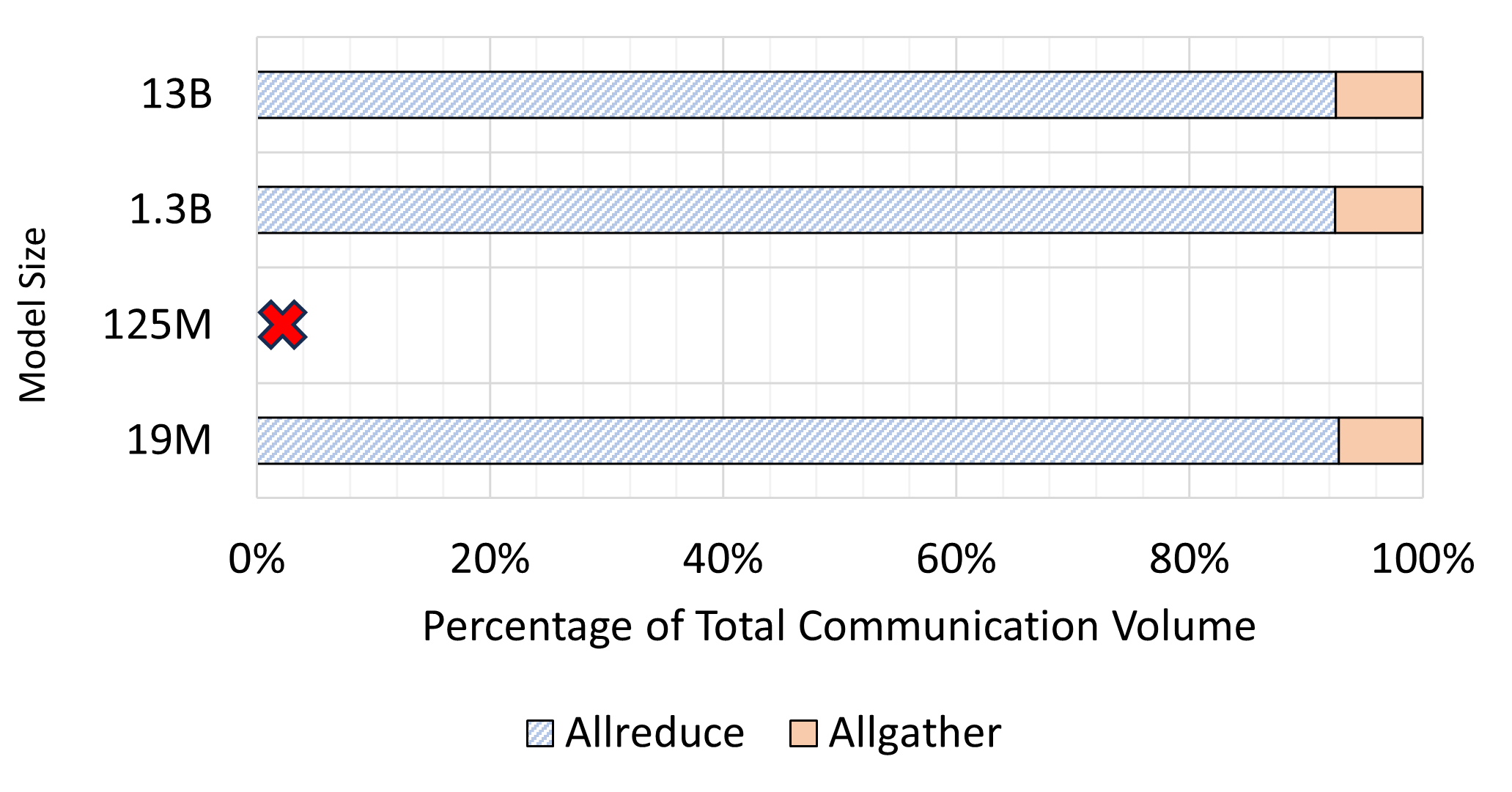}
    \caption{Tensor Parallelism}
\end{subfigure}
\vspace{-1ex}
\caption{Tensor and Pipeline Parallel Communication Breakdown for our four selected model sizes.}
\label{fig:TP-PP_Volume-Perc}
\end{figure*}

Figures \ref{fig:TP-PP_Volume-Perc} shows how differing levels of tensor and pipeline parallelism can affect communication volume\footnote{We saw large Allreduce operations show up in the pure pipeline parallelism case that we suspect are internal to the DeepSpeed framework rather than inherent to the parallelism scheme}. The first immediate observation is the domination of Allgather operations despite the use of point-to-point operations in any configuration utilizing a mix of pipeline and tensor parallelism. Only pure pipeline parallelism avoids this with the next-largest bottleneck being calls to Allreduce\footnote{We saw a larger communication volume than predicted for tensor parallelism, which we believe to be due to DeepSpeed internals}.

\begin{figure*}[ht!]
    \centering
    \begin{subfigure}[t]{0.35\textwidth}
    
    \includegraphics[width=\linewidth]{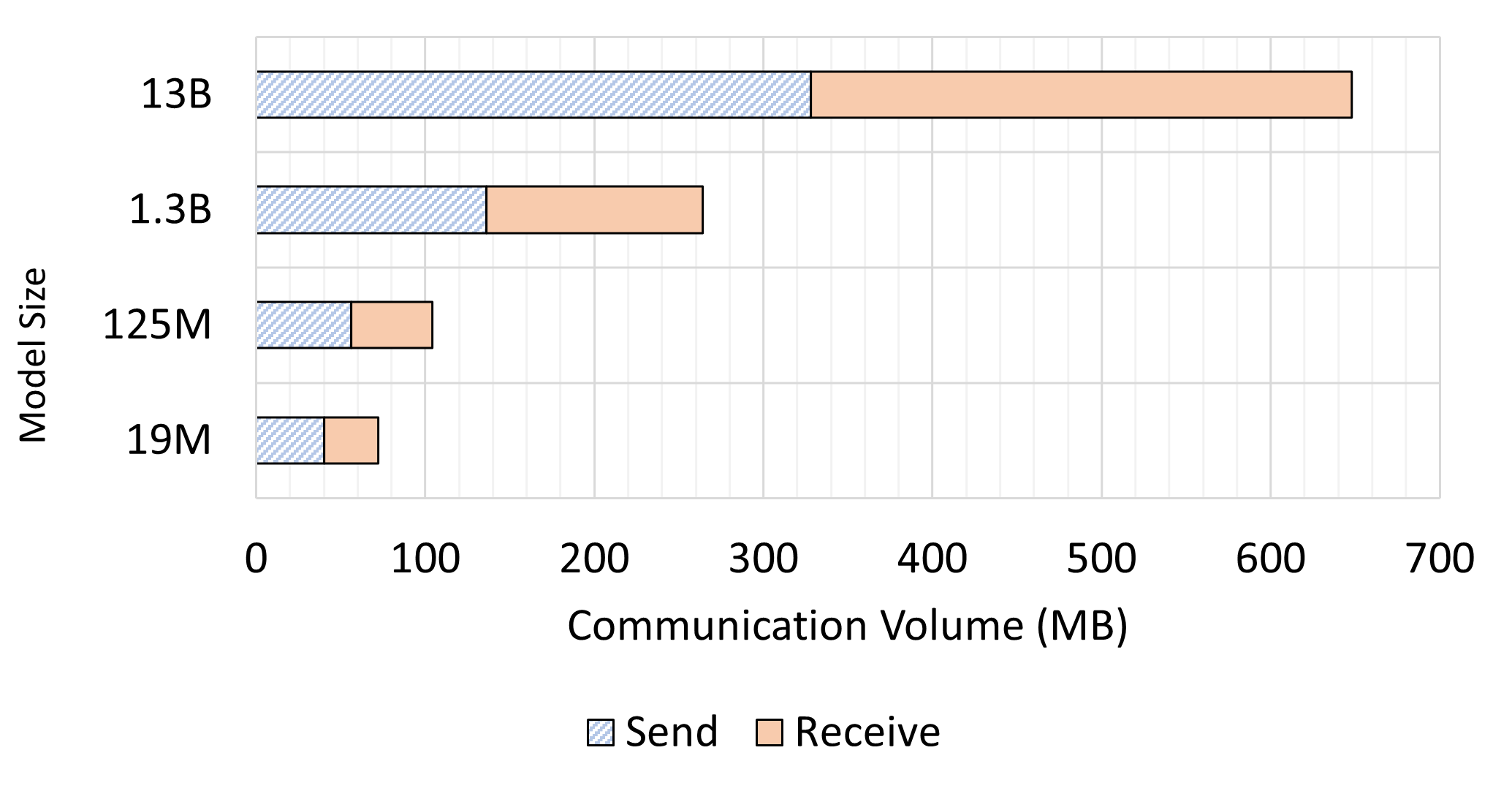}
    \caption{Pipeline Parallelism}
\end{subfigure}
\begin{subfigure}[t]{0.35\textwidth}
    
    \includegraphics[width=\linewidth]{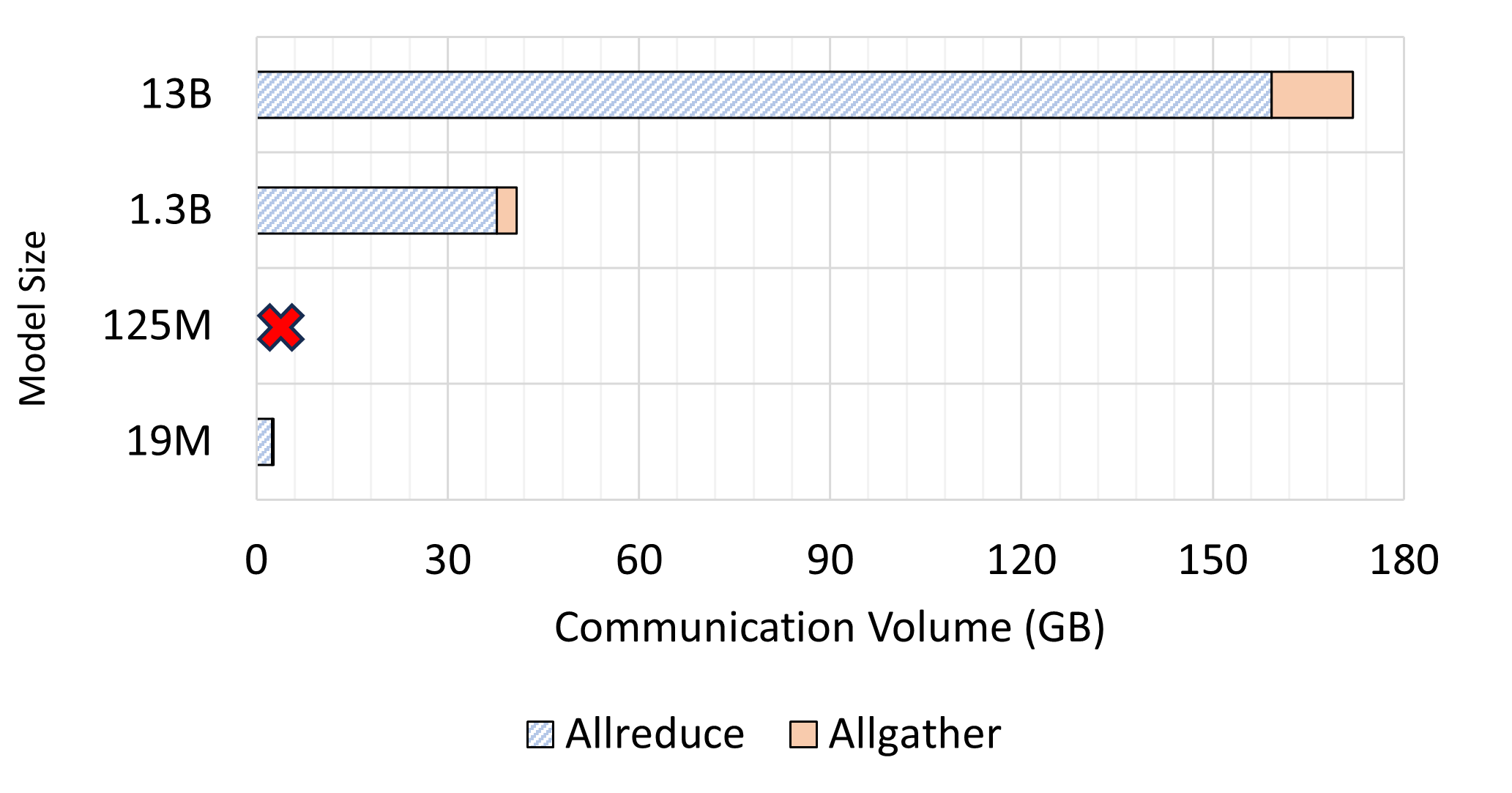}
    \caption{Tensor Parallelism}
\end{subfigure}
\vspace{-1ex}
\caption{Tensor and Pipeline Parallel total communication volume for our four selected model sizes.}
\label{fig:TP-PP_Breakdown}
\end{figure*}

\begin{figure*}[ht!]
    \centering
    \begin{subfigure}[t]{0.35\textwidth}
    
    \includegraphics[width=\linewidth]{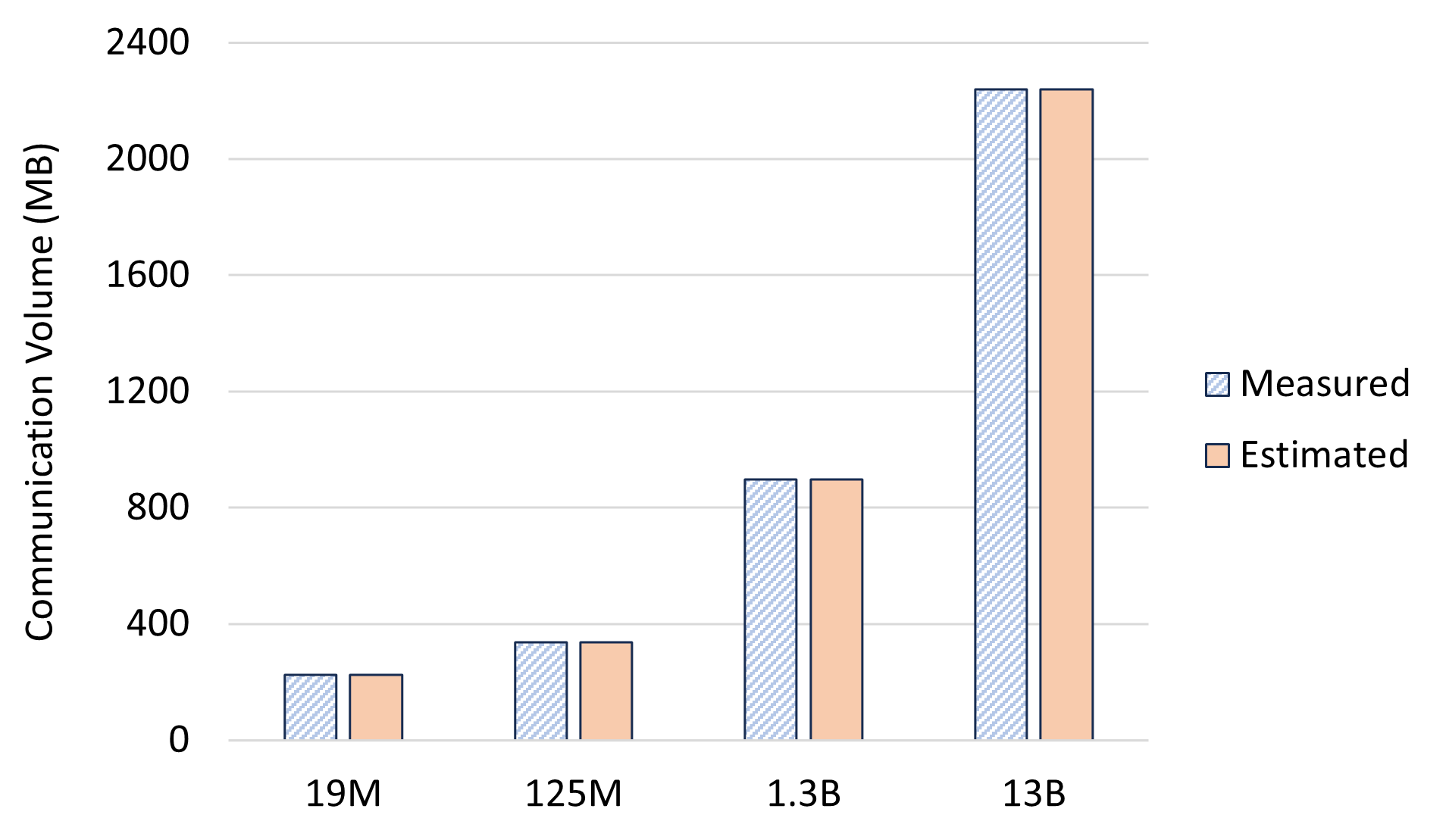}
    \caption{Pipeline Parallelism}
\end{subfigure}
\begin{subfigure}[t]{0.35\textwidth}
    
    \includegraphics[width=\linewidth]{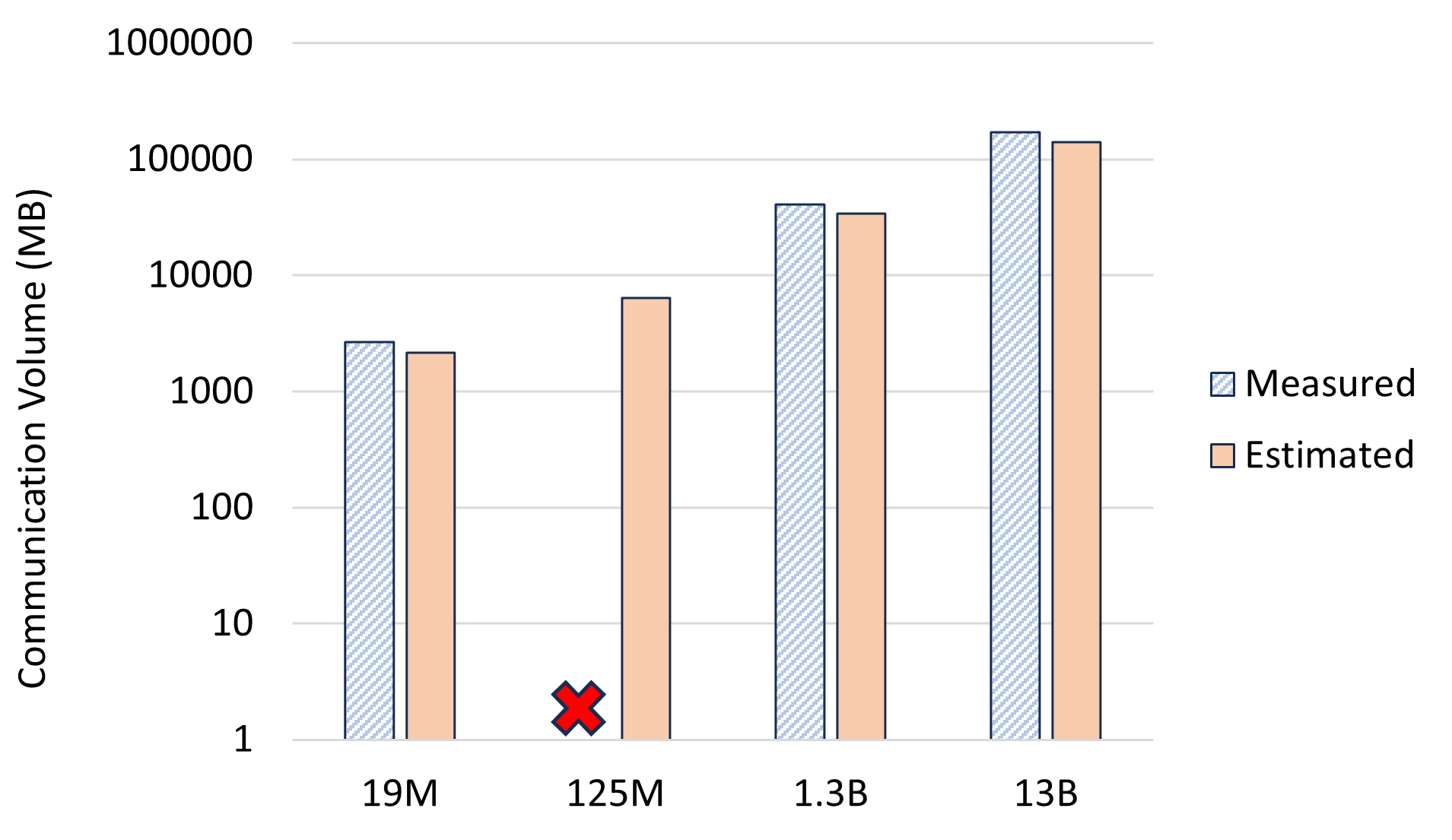}
    \caption{Tensor Parallelism}
\end{subfigure}
\vspace{-1ex}
\caption{Tensor and Pipeline Parallel Communication comparison to theory for our four selected model sizes.}
\label{fig:TP-PP_Theory}
\end{figure*}

Returning to the figures in Section \ref{sec:motivation} we noted that pipeline parallelism has an interesting anomaly: the receive operation is the \textit{only} one to suffer from cold-cache performance, particularly in small message sizes (first iteration receive operations can cause on overhead on the order of thousands of milliseconds). While raw performance modeling is outside the scope of this paper, it is important to note that this anomaly becomes a concern as model size increases and pipeline parallelism is used. This goes back to the takeaway at the end of the previous subsection: Small message optimization is as important as large message optimization.

\subsubsection{Comparison to Performance Model}
Figure\footnote{We note that send operations contain up to an extra eight megabytes. We believe this to be extra metadata being transferred on behalf of the sender} \ref{fig:TP-PP_Theory}\footnote{We note that the 125M-parameter model fails to run with pure tensor parallelism due to the number of attention heads not being appropriately divisible by the number of tensor stages.} shows how the 19M, 125M, 1.3B, and 13B-parameter models perform and match up to the predicted communication volumes based on the Tensor and Pipeline Parallelism formulas from Section \ref{sec:performance-model}. Here, we are primarily interested in the send/receive volume (pipeline parallelism-related) and/or Allreduce communication (tensor parallelism).

\subsection{Sequence Length Experiments}
\label{subsec:SeqLen}

This section examines how sequence length impacts communication behavior for Data-Parallel and Model-Parallel environments. Experiments here were all run on 2 Nodes, 8 GCDs/Node with the 1.3B-parameter model.

\begin{figure*}[ht!]
    \centering
\begin{subfigure}[t]{0.3\textwidth}    \includegraphics[width=\linewidth]{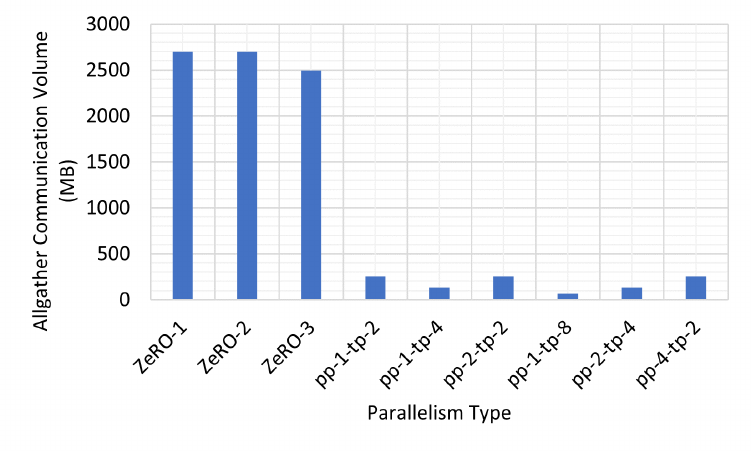}
        \vspace{-1ex}
    \caption{Allgather Comm Volumes for Data/Model Parallelism Schemes}
    \label{fig:Allgather-Seq-Comms}
\end{subfigure}
\begin{subfigure}[t]{0.3\textwidth}  
    \includegraphics[width=\linewidth]{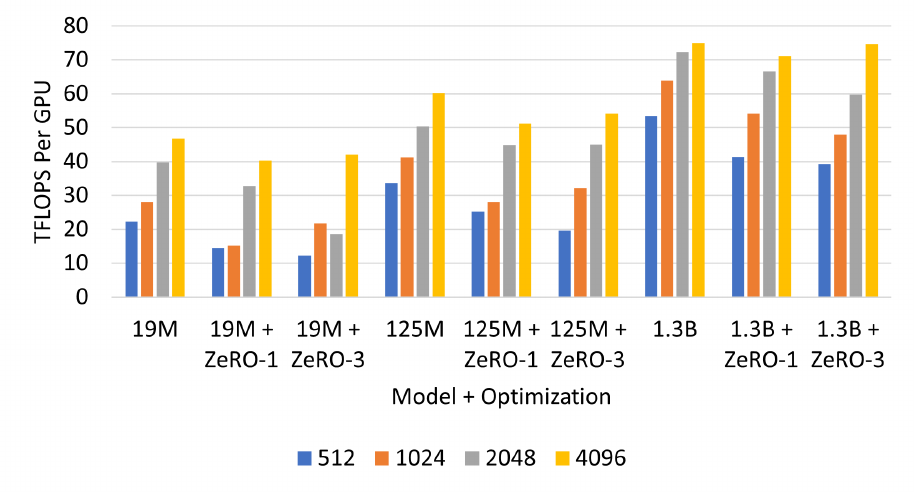}
        \vspace{-1ex}
    \caption{How Sequence Length Impacts Data Parallelism Throughput}
    \label{fig:DP-Throughput}
\end{subfigure} 

\vspace{-1ex}
\caption{Sequence Length Impacts on Allgather, Allreduce, and DP and ZeRO-based throughput}
\label{fig:Seq_Comms_Pt_1}
\end{figure*}

Figure \ref{fig:Allgather-Seq-Comms} shows the Allgather communication volume (where applicable) for both data and model parallelism. To reduce redundancy, we will note that this does not change across increasing sequence length values, from 512 to 4096 or higher. However, we do note that optimizations and sequence length do have an impact on throughput. Figure \ref{fig:DP-Throughput} shows how different levels of ZeRO impact throughput. While we see an approximate 2-2.5x increase in TFlops per GPU, ZeRO optimizations will more often than not result in a decrease of flops for the given sequence length.

\begin{figure*}[ht!]
    \centering
    \begin{subfigure}[t]{0.26\textwidth}
    \includegraphics[width=\linewidth]{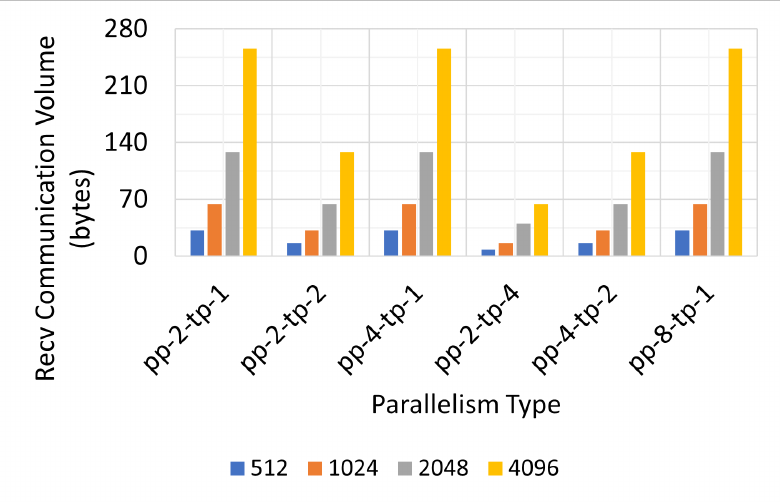}
\vspace{-1ex}
    \caption{Sequence Length Study: Tensor/Pipeline Parallelism Recv Volumes}
    \label{fig:SeqLen-Recv}
\end{subfigure}
\hspace{0.2ex}
\begin{subfigure}[t]{0.24\textwidth}
    \includegraphics[width=\linewidth]{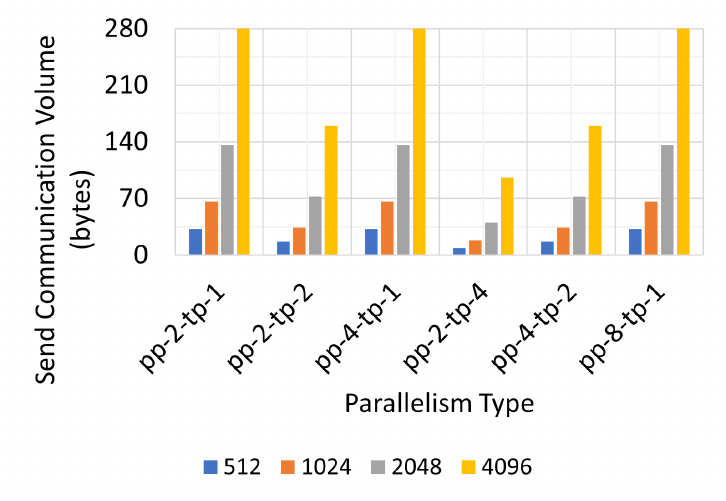}
        \vspace{-1ex}
    \caption{Sequence Length Study: Tensor/Pipeline Parallelism Send Volume}
    \label{fig:SeqLength-Send}
\end{subfigure}
\hspace{0.2ex}
\begin{subfigure}[t]{0.31\textwidth}  
    
    \includegraphics[width=\linewidth]{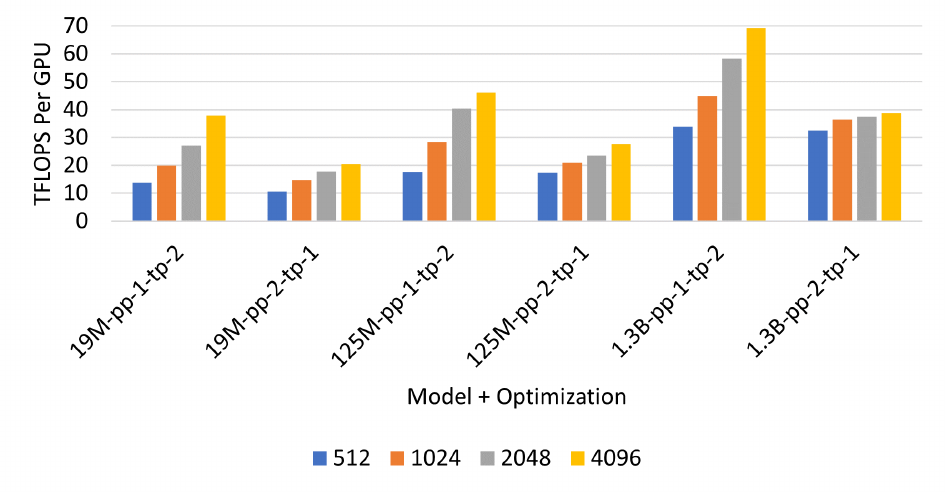}
\vspace{-1ex}
    \caption{How Sequence Length Impacts Tensor and Pipeline Parallelism Throughput}
    \label{fig:SeqLength_MP_TP}
\end{subfigure} 

\vspace{-1ex}
\caption{Sequence Length Impacts on Send/Recv Communication (Communication Volume and Throuhgput)}
\label{fig:Seq_Comms_Pt_2}
\end{figure*}

Compared to data parallelism and ZeRO, there is more variation in the ``key" components tensor/pipeline/model parallelism. While pure tensor parallelism makes sole use of Allreduce, pure pipeline parallelism and model parallelism make use of point-to-point operations as well, and contrary to the above, these volumes increase with token size (see Sections \ref{sec:performance-model} and \ref{subsec:ModelPar}). Figure \ref{fig:SeqLength-Send} shows an approximate doubling/slightly-larger-than-2x increase in communication volume with increasing sequence-length values while Figure \ref{fig:SeqLen-Recv} directly shows a 2x increase with increasing sequence-length values. Similar to the data-parallel results, we also see an increase in throughput as shown in figure \ref{fig:SeqLength_MP_TP}. For brevity, we only show when we have two pipeline stages or a tensor parallelism value of two. Ultimately, the use of tensor parallelism will allow for a higher TFLOP-per-GPU count over pipeline parallelism (up to almost 2x more), though this has an inverse relationship with point-to-point communication (where applicable as pure tensor parallelism does not use point-to-point) in communication volume.

\section{Related Work}
\label{sec:related}
\vspace{-1ex}
Many papers have analyzed LLMs and characterized them through bias and truthfulness. The authors of \cite{Caricature-LLM-Sim} develop ``CoMPosT" to characterize LLM simulations that result in caricatures: misrepresentations of the models/workloads being simulated. Our work performs analysis at a system level to show the impact of communication on these models. \cite{NEURIPS2023_ae9500c4} focuses on LLMs as a data generator and characterizes the diversity and bias of the data it generates post-training.

Research has been done to characterize the performance of DNNs on HPC clusters. \cite{Awan-DNN-Char} and \cite{Jain-DNN-Char-PyTorch} characterized DNN performance, first in the context of CPU/GPU-based architectures and later with the PyTorch and TensorFlow frameworks. The authors of \cite{Awan-CUDA-Aware-MPI-Char} evaluated DNN performance in the context of CUDA-aware MPI libraries. 

More recently, LLMs have been analyzed from a system/performance perspective. The authors of \cite{LLM-Arch} analyze different LLM architectures on the current\footnote{As of May 2024, Frontier ranks first in the Top500 list with an Rpeak of 1.7 exaFLOPS.} world's fastest supercomputer Frontier and answer the question of how different model architectures impact performance. The authors of \cite{LLM-datacenter} explored the impact of LLMs on large-scale systems, namely hardware limitations and capabilities. They note communication overheads as part of some performance skew and degradation but ultimately do not do in-depth communication analysis. Even more recently, the authors of \cite{MegaScale} designed, developed, and characterized the performance of their ``MegaScale" framework to allow for easy training/deployment of LLMs for scales at and beyond ten thousand GPUs, with a focus on software/hardware co-design for efficiency and stability. A more recent work (\cite{LLM_Scale_out_Char}) looks at characterizing LLM performance at scale on NVIDIA DGX clusters with an emphasis on 200Gb/s network utilization. Their work differs from ours in that they look at performance characterization concerning scale, not directly in communication volume and behavior. They also do not evaluate model, tensor, or pipeline parallelism and how a combination of sequence length and parallelism scheme impacts communication volume and throughput.
\section{Conclusions}
\label{sec:conclusions}

We have presented a characterization of LLM communication behavior on the Frontier supercomputer. This has been done by combining a rigorous performance model for multiple parallelism schemes and multiple experiments utilizing current state-of-the-art training frameworks with precise profiling of communication and compute. We have provided insights into potential optimizations for communication middleware for small-message communication. For future pending work, given that the Frontier system represents one combination, we would like to examine further parallelism schemes here such as multi-dimensional parallelism and expert parallelism. We would also like to examine how all the schemes presented here might change on current and upcoming systems with new or maturing communication and software stacks such as Aurora at Argonne National Lab (Intel GPUs and Intel CPUs) or the upcoming Vista cluster at the Texas Advanced Computing Center (NVIDIA Grace Hopper).

\section{Acknowledgments}
 We would like to thank the Oak Ridge Computing Facilities/Oak Ridge National Laboratory for granting us access to the Frontier supercomputer to run our experiments. This research is supported in part by NSF grants \#1818253, \#1854828, \#2007991, \#2018627, \#2112606, \#2311830, \#2312927, and XRAC grant \#NCR-130002.

\bibliographystyle{IEEEtran}
\bibliography{dl_characterization.bib}

\begin{thebibliography}{10}
\providecommand{\url}[1]{#1}
\csname url@samestyle\endcsname
\providecommand{\newblock}{\relax}
\providecommand{\bibinfo}[2]{#2}
\providecommand{\BIBentrySTDinterwordspacing}{\spaceskip=0pt\relax}
\providecommand{\BIBentryALTinterwordstretchfactor}{4}
\providecommand{\BIBentryALTinterwordspacing}{\spaceskip=\fontdimen2\font plus
\BIBentryALTinterwordstretchfactor\fontdimen3\font minus \fontdimen4\font\relax}
\providecommand{\BIBforeignlanguage}[2]{{%
\expandafter\ifx\csname l@#1\endcsname\relax
\typeout{** WARNING: IEEEtran.bst: No hyphenation pattern has been}%
\typeout{** loaded for the language `#1'. Using the pattern for}%
\typeout{** the default language instead.}%
\else
\language=\csname l@#1\endcsname
\fi
#2}}
\providecommand{\BIBdecl}{\relax}
\BIBdecl

\bibitem{openai2024gpt4}
{OpenAI and Josh Achiam and Steven Adler and Sandhini Agarwal and Lama Ahmad and Ilge Akkaya and et. al.}, ``{GPT-4 Technical Report},'' 2024.

\bibitem{geminiteam2024gemini}
{Gemini Team and Rohan Anil and Sebastian Borgeaud and Jean-Baptiste Alayrac and Jiahui Yu and Radu Soricut and et. al.}, ``{Gemini: A Family of Highly Capable Multimodal Models},'' 2024.

\bibitem{touvron2023llama}
H.~Touvron, L.~Martin, K.~Stone, P.~Albert, A.~Almahairi, Y.~Babaei, and et. al., ``{Llama 2: Open Foundation and Fine-Tuned Chat Models},'' 2023.

\bibitem{PaLM24}
A.~Chowdhery, S.~Narang, J.~Devlin, M.~Bosma, G.~Mishra, A.~Roberts, and et. al., ``{PaLM: scaling language modeling with pathways},'' \emph{J. Mach. Learn. Res.}, vol.~24, no.~1, mar 2024.

\bibitem{MegaScale}
\BIBentryALTinterwordspacing
Z.~Jiang, H.~Lin, Y.~Zhong, Q.~Huang, Y.~Chen, Z.~Zhang, and et. al., ``{{MegaScale}: Scaling Large Language Model Training to More Than 10,000 {GPUs}},'' in \emph{21st USENIX Symposium on Networked Systems Design and Implementation (NSDI 24)}.\hskip 1em plus 0.5em minus 0.4em\relax Santa Clara, CA: USENIX Association, Apr. 2024, pp. 745--760. [Online]. Available: \url{https://www.usenix.org/conference/nsdi24/presentation/jiang-ziheng}
\BIBentrySTDinterwordspacing

\bibitem{megatron-lm}
NVIDIA, ``{Megatron-LM: Ongoing research training transformer models at scale},'' \url{https://github.com/NVIDIA/Megatron-LM}, 2024, accessed: \today.

\bibitem{meta-llama}
Meta, ``{Meta Llama},'' \url{https://llama.meta.com/}, 2024, accessed: \today.

\bibitem{deepspeed-mii}
``{DeepSpeed-MII},'' \url{https://github.com/microsoft/DeepSpeed-MII}, 2022.

\bibitem{msccl}
Microsoft, ``{MSCCL: Microsoft Collective Communication Library},'' \url{https://github.com/microsoft/msccl}, 2024, accessed: \today.

\bibitem{rccl}
ROCm, ``{RCCL: ROCm Communication Collectives Library},'' \url{https://github.com/ROCm/rccl}, 2024, accessed: \today.

\bibitem{nccl}
NVIDIA, ``{NVIDIA Collective Communications Library (NCCL)},'' \url{https://developer.nvidia.com/nccl}, 2024, accessed: \today.

\bibitem{mcr-dl}
\BIBentryALTinterwordspacing
Q.~Anthony, A.~Awan, J.~Rasley, Y.~He, A.~Shafi, M.~Abduljabbar, H.~Subramoni, and D.~Panda, ``{MCR-DL: Mix-and-Match Communication Runtime for Deep Learning},'' in \emph{2023 IEEE International Parallel and Distributed Processing Symposium (IPDPS)}.\hskip 1em plus 0.5em minus 0.4em\relax Los Alamitos, CA, USA: IEEE Computer Society, may 2023, pp. 996--1006. [Online]. Available: \url{https://doi.ieeecomputersociety.org/10.1109/IPDPS54959.2023.00103}
\BIBentrySTDinterwordspacing

\bibitem{awan_hoti_19}
A.~Jain, A.~A. Awan, C.~Chu, H.~Subramoni, and D.~Panda, ``Communication profiling and characterization of deep-learning workloads on clusters with high-performance interconnects,'' August 2019.

\bibitem{Jain-DNN-Char-PyTorch}
A.~Jain, A.~A. Awan, Q.~Anthony, H.~Subramoni, and D.~K.~D. Panda, ``{Performance Characterization of DNN Training using TensorFlow and PyTorch on Modern Clusters},'' in \emph{2019 IEEE International Conference on Cluster Computing (CLUSTER)}, 2019, pp. 1--11.

\bibitem{gpt-neox-library}
\BIBentryALTinterwordspacing
A.~Andonian, Q.~Anthony, S.~Biderman, S.~Black, P.~Gali, L.~Gao, E.~Hallahan, J.~Levy-Kramer, C.~Leahy, L.~Nestler, K.~Parker, M.~Pieler, J.~Phang, S.~Purohit, H.~Schoelkopf, D.~Stander, T.~Songz, C.~Tigges, B.~Thérien, P.~Wang, and S.~Weinbach, ``{GPT-NeoX}: Large scale autoregressive language modeling in {PyTorch},'' GitHub Repo, 9 2023. [Online]. Available: \url{https://www.github.com/eleutherai/gpt-neox}
\BIBentrySTDinterwordspacing

\bibitem{vaswani2017attention}
A.~Vaswani, N.~Shazeer, N.~Parmar, J.~Uszkoreit, L.~Jones, A.~N. Gomez, {\L}.~Kaiser, and I.~Polosukhin, ``{Attention is All You Need},'' \emph{Advances in Neural Information Processing Systems}, vol.~30, 2017.

\bibitem{devlin2018bert}
J.~Devlin, M.-W. Chang, K.~Lee, and K.~Toutanova, ``{Bert: Pre-training of deep bidirectional transformers for language understanding},'' \emph{arXiv preprint arXiv:1810.04805}, 2018.

\bibitem{radford2019language}
A.~Radford, J.~Wu, R.~Child, D.~Luan, D.~Amodei, I.~Sutskever \emph{et~al.}, ``Language models are unsupervised multitask learners,'' \emph{OpenAI blog}, vol.~1, no.~8, p.~9, 2019.

\bibitem{korthikanti2022reducing}
V.~Korthikanti, J.~Casper, S.~Lym, L.~McAfee, M.~Andersch, M.~Shoeybi, and B.~Catanzaro, ``Reducing activation recomputation in large transformer models,'' 2022.

\bibitem{hoffmann2022training}
J.~Hoffmann, S.~Borgeaud, A.~Mensch, E.~Buchatskaya, T.~Cai, E.~Rutherford, D.~de~Las~Casas, L.~A. Hendricks, J.~Welbl, A.~Clark, T.~Hennigan, E.~Noland, K.~Millican, G.~van~den Driessche, B.~Damoc, A.~Guy, S.~Osindero, K.~Simonyan, E.~Elsen, J.~W. Rae, O.~Vinyals, and L.~Sifre, ``Training compute-optimal large language models,'' 2022.

\bibitem{kaplan2020scaling}
J.~Kaplan, S.~McCandlish, T.~Henighan, T.~B. Brown, B.~Chess, R.~Child, S.~Gray, A.~Radford, J.~Wu, and D.~Amodei, ``Scaling laws for neural language models,'' 2020.

\bibitem{bennun2018demystifying}
T.~Ben-Nun and T.~Hoefler, ``{Demystifying Parallel and Distributed Deep Learning: An In-Depth Concurrency Analysis},'' 2018.

\bibitem{you2020large}
Y.~You, J.~Li, S.~Reddi, J.~Hseu, S.~Kumar, S.~Bhojanapalli, X.~Song, J.~Demmel, K.~Keutzer, and C.-J. Hsieh, ``{Large Batch Optimization for Deep Learning: Training BERT in 76 minutes},'' 2020.

\bibitem{huang2019gpipe}
Y.~Huang, Y.~Cheng, A.~Bapna, O.~Firat, M.~X. Chen, D.~Chen, H.~Lee, J.~Ngiam, Q.~V. Le, Y.~Wu, and Z.~Chen, ``{GPipe: Efficient Training of Giant Neural Networks using Pipeline Parallelism},'' 2019.

\bibitem{harlap2018pipedream}
A.~Harlap, D.~Narayanan, A.~Phanishayee, V.~Seshadri, N.~Devanur, G.~Ganger, and P.~Gibbons, ``{PipeDream: Fast and Efficient Pipeline Parallel DNN Training},'' 2018.

\bibitem{shoeybi2019megatron}
M.~Shoeybi, M.~Patwary, R.~Puri, P.~LeGresley, J.~Casper, and B.~Catanzaro, ``Megatron-{LM}: {T}raining {M}ulti-{B}illion {P}arameter {L}anguage {M}odels using {GPU} {M}odel {P}arallelism,'' \emph{arXiv preprint arXiv:1909.08053}, 2019.

\bibitem{LLM_Scale_out_Char}
\BIBentryALTinterwordspacing
S.~Cheng, J.-L. Lin, M.~Emani, S.~Raskar, S.~Foreman, Z.~Xie, V.~Vishwanath, and M.~T. Kandemir, ``Thorough characterization and analysis of large transformer model training at-scale,'' \emph{Proc. ACM Meas. Anal. Comput. Syst.}, vol.~8, no.~1, feb 2024. [Online]. Available: \url{https://doi.org/10.1145/3639034}
\BIBentrySTDinterwordspacing

\bibitem{ZeRO}
S.~Rajbhandari, J.~Rasley, O.~Ruwase, and Y.~He, ``{ZeRO: Memory Optimizations Toward Training Trillion Parameter Models},'' 2020.

\bibitem{dettmers20228bit}
T.~Dettmers, M.~Lewis, S.~Shleifer, and L.~Zettlemoyer, ``8-bit optimizers via block-wise quantization,'' 2022.

\bibitem{wang2023zero}
G.~Wang, H.~Qin, S.~A. Jacobs, C.~Holmes, S.~Rajbhandari, O.~Ruwase, F.~Yan, L.~Yang, and Y.~He, ``Zero++: Extremely efficient collective communication for giant model training,'' 2023.

\bibitem{LLM-Arch}
J.~Yin, A.~Bose, G.~Cong, I.~Jyngaas, and Q.~Anthony, ``{Comparative Study of Large Language Model Architectures on Frontier },'' 5 2024, accepted, to be presented at IPDPS 2024.

\bibitem{Caricature-LLM-Sim}
\BIBentryALTinterwordspacing
M.~Cheng, T.~Piccardi, and D.~Yang, ``{{C}o{MP}os{T}: Characterizing and Evaluating Caricature in {LLM} Simulations},'' in \emph{Proceedings of the 2023 Conference on Empirical Methods in Natural Language Processing}, H.~Bouamor, J.~Pino, and K.~Bali, Eds.\hskip 1em plus 0.5em minus 0.4em\relax Singapore: Association for Computational Linguistics, 12 2023, pp. 10\,853--10\,875. [Online]. Available: \url{https://aclanthology.org/2023.emnlp-main.669}
\BIBentrySTDinterwordspacing

\bibitem{NEURIPS2023_ae9500c4}
\BIBentryALTinterwordspacing
Y.~Yu, Y.~Zhuang, J.~Zhang, Y.~Meng, A.~J. Ratner, R.~Krishna, J.~Shen, and C.~Zhang, ``{Large Language Model as Attributed Training Data Generator: A Tale of Diversity and Bias},'' in \emph{{Advances in Neural Information Processing Systems}}, A.~Oh, T.~Naumann, A.~Globerson, K.~Saenko, M.~Hardt, and S.~Levine, Eds., vol.~36.\hskip 1em plus 0.5em minus 0.4em\relax Curran Associates, Inc., 2023, pp. 55\,734--55\,784. [Online]. Available: \url{https://proceedings.neurips.cc/paper\_files/paper/2023/file/ae9500c4f560\\7caf2eff033c67daa9d7-Paper-Datasets\_and\_Benchmarks.pdf}
\BIBentrySTDinterwordspacing

\bibitem{Awan-DNN-Char}
\BIBentryALTinterwordspacing
A.~A. Awan, H.~Subramoni, and D.~K. Panda, ``{An In-depth Performance Characterization of CPU- and GPU-based DNN Training on Modern Architectures},'' in \emph{Proceedings of the Machine Learning on HPC Environments}, ser. MLHPC'17.\hskip 1em plus 0.5em minus 0.4em\relax New York, NY, USA: Association for Computing Machinery, 2017. [Online]. Available: \url{https://doi.org/10.1145/3146347.3146356}
\BIBentrySTDinterwordspacing

\bibitem{Awan-CUDA-Aware-MPI-Char}
A.~A. Awan, J.~Bédorf, C.-H. Chu, H.~Subramoni, and D.~K. Panda, ``{Scalable Distributed DNN Training using TensorFlow and CUDA-Aware MPI: Characterization, Designs, and Performance Evaluation},'' in \emph{2019 19th IEEE/ACM International Symposium on Cluster, Cloud and Grid Computing (CCGRID)}, 2019, pp. 498--507.

\bibitem{LLM-datacenter}
\BIBentryALTinterwordspacing
Q.~Hu, Z.~Ye, Z.~Wang, G.~Wang, M.~Zhang, Q.~Chen, P.~Sun, D.~Lin, X.~Wang, Y.~Luo, Y.~Wen, and T.~Zhang, ``{Characterization of Large Language Model Development in the Datacenter},'' in \emph{21st USENIX Symposium on Networked Systems Design and Implementation (NSDI 24)}.\hskip 1em plus 0.5em minus 0.4em\relax Santa Clara, CA: USENIX Association, 4 2024, pp. 709--729. [Online]. Available: \url{https://www.usenix.org/conference/nsdi24/presentation/hu}
\BIBentrySTDinterwordspacing

\end{thebibliography}

\end{document}